# Vacuum Dealloyed Brass as Li-Metal Battery Current Collector: Effect of Zinc and Porosity

Eric V. Woods[1,*], Xinren Chen[1], Shaolou Wei[1], Yuwei Zhang[1], Alisson Kwiatkowski da Silva[1], Ayman A. El-Zoka[1,2], J. Manoj Prabhakar[1], Tim M. Schwarz[1], Yongqiang Kang[1], Leonardo S. Aota[1], Mahander P. Singh[1], Katja Angenendt[1], Özge Özgün[1], Matic Jovičević-Klug[1], Patricia Jovičević-Klug[1], Christian Broß[1], Jian Liu[3], René de Kloe[4], Gerhard Dehm[1], Stefan Zaefferer[1], Yug Joshi[1,*], Baptiste Gault[1,2,*]

1. Max-Planck-Institute for Sustainable Materials GmbH (formerly Max-Planck-Institut für Eisenforschung GmbH), 40237 Düsseldorf, Germany
2. Department of Materials, Royal School of Mines, Imperial College London, SW7 2AZ, London, UK.
3. School of Engineering, Faculty of Applied Science, The University of British Columbia, Kelowna V1V 1V7, Canada
4. EDAX / Gatan, Ringbaan Noord 103, 5046AA Tilburg, The Netherlands

*Corresponding Authors: e.woods@mpie.de, y.joshi@mpie.de, b.gault@mpie.de

## Abstract

"Anode-free" lithium-metal batteries promise significantly higher energy density than conventional graphite-based lithium-ion batteries; however, lithium dendrite growth can lead to internal short circuits with associated safety risks. While porous current collectors can suppress dendrite growth, optimal porosity and composition remain unknown. Here, we show that the temperature during vapor phase dealloying (VPD) of α-brass ($Cu_{63}Zn_{37}$) controls the surface Zn concentration, decreasing from 8% to below 1% from 500 – 800°C. The surface composition is controlled by the temperature-dependent diffusion. A battery cell maintains > 90% Coulombic efficiency (CE) over 100 cycles when the Zn content is the lowest, whereas the higher-Zn samples degraded to ~70% CE. The difference in surface composition has hence dramatic effects on battery performance, and our results demonstrate how precise compositional control enables stable Li-metal battery operation, establishing ~1 at.% surface Zn as optimal for preventing capacity fading and uniform lithium plating, while establishing predictive relationships between processing temperature and surface composition. This work provides design rules for multifunctional current collectors and demonstrates scalable VPD production for next-generation batteries.

## Introduction

Li-ion batteries (LIBs) using Li metal anodes theoretically offer an order of magnitude improvement in specific capacity vs. current graphite anodes (3,860 mAh·g$^{-1}$ vs 372 mAh·g$^{-1}$) [1,2]. That advance would enable both substantially increased electric vehicle range and reduced battery weight, as well as long-duration stationary [1,3]. "Anode-free" Li-metal batteries, which begin with a bare current collector with no Li metal and plate Li directly onto the current collector, offer further simplified cell design and safety [4–6]. However, uncontrolled Li dendrite formation at local "hot spots" – higher local current – during Li plating and stripping remains the greatest challenge, since these can lead to internal short circuits with safety risks including catastrophic thermal runaway including fires [7–10]. Porous metal current collectors suppress Li dendrite formation by reducing local current density through increasing surface area, providing controlled nucleation sites, and improved Li ion transport, which all improve Li plating uniformity [11–15].

Porous metals can be produced through diverse fabrication techniques like gas injection into melts, powder metal sintering, templating methods, etc. [16–24], and find wide uses from catalysis to structural weight reduction [25–28]. For battery current collectors needing both high thermal and electrical conductivity and controlled porosity, dealloying offers distinct advantages. The most relevant systems use the more "noble" metal (e.g. chemically stable) Cu [29–31] produced from Cu binary alloys [29,32–34] formed with a less noble, sacrificial metal like Al [35], Zn [36], or Mn [37]. Au-based systems, while less relevant for battery applications due to cost, normally use Ag [38] as the sacrificial metal and have contributed to chemical dealloying fundamentals. These alloys can be deposited as thin films [39,40] or prepared from bulk; example systems are Au-Ag [40–42] and Cu-Zn [43–45].

Chemical or electrochemical "wet" dealloying methods are predominantly used for dealloying Au and Cu, which etch the sacrificial metal with strong acids like HCl [46] or HNO$_3$ [41] or strong bases like NaOH [34,47]. Those processes produce significant chemical waste and the sacrificial metal is dissolved and difficult to recover, and often fail to completely remove Zn from nanoporous Cu [43]. In contrast, vapor phase dealloying (VPD), also referred to as vacuum dealloying, can produce porous Cu with negligible remaining Zn in one step and potentially easily recover the sacrificial metal, as illustrated schematically in **Figure 1(a)**. The VPD process exploits the high vapor pressure and lower melting point of Zn, causing preferential evaporation from Cu-Zn binary alloys [43,48–51].

Successful VPD requires Zn content above the parting limit – approximately 20 at.%, the minimum Zn composition required for nanopore formation [52]. Optimal dealloying conditions to reach a certain composition remain unclear. Above this threshold, the vapor pressure difference between Cu and Zn creates a strong thermodynamic driving force for Zn removal. As Zn atoms preferentially evaporate from the surface, there is a driving force for Zn atoms to diffuse towards the surface to replace them. This is accompanied by vacancy formation in the crystal lattice, that gradually coalesce together into interconnected pores, a process known as the Kirkendall effect [53–57].

VPD using Cu-Zn only a few percent above the parting limit has shown substantial nanopore formation even in very thin sheets, e.g. 20 µm [11]. However, most VPD fabrication has used higher Zn compositions, e.g. up to 70 at.% Zn, to achieve rapid pore formation [32,46,58–60]. While higher sacrificial metal content enables easier removal and higher consistent porosity, commercially available α-brass (37 at.% Zn) closer to the parting limit provide distinct advantages for reliable pore formation while retaining mechanical integrity. Since nanoporous Cu formed from high-Zn alloys is structurally brittle, practical applications have used bilayer foils with a normal Cu backing layer for mechanical stability [59]. Typically, only small pieces are produced because they are fragile [29,61]. Analysis of the Cu-Zn phase diagram indicates that processing temperatures between 500 °C and 800 °C are optimal to avoid any liquidus or shifting to ß-phase [62,63].

Porous Cu current collectors have been tested for Li-ion batteries before and demonstrated improved Li plating and performance compared to standard flat Cu foil, thereby suppressing dendrite growth even at high current density [14,64–68], as reviewed in Ref. [69] for instance. Specifically, VPD-fabricated porous Cu current collectors were typically thin sheets with a 10 – 50 µm thickness [11]. However, significant differences in battery performance metrics, i.e. Coulombic efficiency (CE), defined as the ratio of discharge capacity over charged capacity in one battery cycle, and capacity retention after cycling, have been reported when comparing VPD-prepared Cu foil when compared to Cu foam [70]. These inconsistent results indicate that specific VPD processing parameters strongly influence battery performance, yet systematic studies of VPD processing parameters are rare, especially for battery-relevant scales.

When Zn or Zn-based alloys were used as the anode in organic electrolyte-based LIBs, they suppressed Li dendrite growth [26,71]. Recently, dual-metal alloy current collectors demonstrated that Zn alloys with Li and can be beneficial for LIB performance [72]. Additionally, recent work using flat Cu-Zn foil as an anode or current collector in Li-ion batteries, demonstrated good success in suppressing Li dendrite growth [73,74]. Moreover, previous work suggested that Cu-Zn could function better than pure Zn-based anodes to increase CE and to provide enhanced nucleation for more uniform Li plating [75]. Likewise, current collectors or anodes which alloy with Li have proven promising to suppress dendrites and improve plating, like In, Ag, Sn, Zn, and other metals [76–79].

Given the benefits of Zn alloying, VPD can reliably produce final surface Zn levels from below 1 at.% to the maximum concentration in the original alloy. Testing current collectors with different Zn compositions prepared at different temperatures in anode-free LIBs simultaneously enables evaluation of both the effects of differing Zn surface concentrations as well as porosity to Li plating and battery performance.

Despite prior work on both porous Cu and Zn-containing current collectors [47,74,80], no study has systematically examined how VPD temperature simultaneously changes both porosity and Zn concentration for lower Zn alloys, e.g. α-brass. In this work, the effects of temperature on Zn removal from VPD-prepared brass under high vacuum at both the surface and in the bulk as well as porosity are examined, tracking compositional changes at the surface and throughout the 0.4 mm thickness. An array of advanced characterization techniques is used to provide compositional and structural information across varying length scales, from μm to nm. The resultant compositions and gradients are compared with thermodynamic simulations to understand the underlying mechanisms, which helps guide optimization of porous Cu and alloy preparation. This work addresses that previous research gap by examining VPD temperature effects. Subsequently, the porous Cu-Zn formed at different temperatures is tested in Li metal battery half cells to evaluate different Zn surface concentrations. Overall, the optimal Zn composition for suppressing dendrites in porous Cu-Zn while still functioning as an effective current collector is determined. VPD provides a quickly scalable, affordable one-step process using a common bulk commercial alloy with the possibility to recover all sacrificial metal to achieve circularity. This study demonstrates that VPD-produced porous Cu-Zn current collectors achieve uniform Li plating by tailoring the Zn surface and pore concentration [72,81], and secondarily by porosity [80], providing practical solutions for suppressing dendrite formation in commercial anode-free LIBs .

## Methods

### Materials

Yellow brass, e.g. α-brass sheets (e.g. Cu: 63% Zn: 37%), hereafter CuZn, were obtained from Metall Ehrnsberger GbR (Metall Ehrnsberger GbR, Teublitz, Germany); thickness was 0.4 mm unless otherwise specified. X-ray Diffraction (XRD) data showed the as-received metal to be single phase α-brass between 5 – 7 μm described in [82].

## Sample Preparation, including Polishing

The received material was cut into 3 mm x 7 mm pieces or 10 mm discs, all 0.4 mm in thickness. The pieces were then annealed under argon at 600 °C for 1 hour with furnace cooling to relieve mechanical damage and stress from cutting, which increased the average grain size from 5 – 7 µm to 12.6 µm, as discussed in [43,81]. VPD was conducted in a custom high vacuum furnace pumped to a pressure of $1\times10^{-5}$ Torr at temperatures in the 600 – 800 °C range, for 1 hour and at 500 °C for 2 hours, both with furnace cooling under vacuum, unless otherwise specified. As can be seen in Fig. 1 (a). The brass discs were polished to 4000 grit before VPD for battery samples.

When required, sample polishing of non-dealloyed pieces was done as in [82] for metallurgical analysis. After polishing, such samples were rinsed with ethanol, dried with Ar, and stored in a desiccator. For purposes of making the battery flat brass cells, the brass discs were ground progressively with 400, 800, 1200, 2000, and 4000 grit polishing pads at 120rpm followed by a short time with 1 µm polishing cloth using diamond suspension, following procedures above.

For dealloyed samples, the porous nature and irregular size of the material required a different strategy for cross-sectional polishing. Using custom Cu-clamps for polishing was unsuccessful. As a result, the 3 mm x 7 mm x 0.4 mm pieces were placed in a plastic clip holder to hold the pieces in a vertical orientation, and then hot mounted with Polyfast thermoplastic resin (Struers ApS, Ballerup, Denmark) for 5 minutes using a Struers CitoPress-15 mounting press. The vertically mounted pieces were then prepared as above, followed by polishing with OPS suspension (0.05 µm colloidal silica) to remove approximately 1 mm of material to ensure that the bulk of the specimen was reached for further Electron Dispersive X-ray Spectroscopy (EDX) analysis.

## X-ray Based Characterization Techniques – X-Ray Diffraction (XRD) and X-ray Photoelectron Spectroscopy (XPS)

XRD data was acquired using different systems: a Bruker Diffractometer D8 Advance A25-X system (Bruker AXS GmbH, Karlsruhe, Germany) and a Rigaku Smartlab 9 kW system (Rigaku Corporation, Tokyo, Japan). Qualitative phase analysis was done with Bruker DEIFFRAC EVA software, version 4.3.0.1. Bruker TOPAS software, version 5.0, was used for Reitveld refinement, with NIST standards for broadening: $Al_2O_3$ + Si, as well as Match 4 software (Crystal Impact, Bonn, Germany).

Note that for 500 °C measurements, the Bruker system was used with a Co Kα X-ray tube; for the 500 °C glancing angle measurements, the Rigaku system with a Cu Kα X-ray tube was used; and for the 600, 700, 750, 800, and 900 °C measurements the Bruker system with a Cu Kα X-ray tube was used.

Additional details on the instruments, settings, X-ray optics, and measurements can be found in **Table T1** and **Table T2.**

Estimated XRD penetration depths can be calculated for these conditions can be approximated using the AbsorbX software package. For Cu Kα radiation analysis of pure Cu substrates, the computed maximum penetration depth is at 2θ = 43° and is approximately 9 μm; for Co Kα radiation, the computed maximum depth occurs for 2θ = 51° and is approximately 7 μm.

XPS data was acquired on a Physical Electronics PHI Quantera II (Physical Electronics, Inc., Changhassen, MN, USA), and analysed with CasaXPS 2.3.15 (Casa Software Ltd., Devon, UK). XPS was carried out on a Physical Electronics PHI Quantera II instrument equipped with an Al-Kα monochromatic X-ray source (1486.6 eV). All spectra were recorded at a 45° take-off angle with a 100 μm X-ray spot size. Survey scans utilized a pass energy of 112 eV and an energy step size of 0.1 eV. For detailed chemical state analysis, high-resolution scans of the O 1s, C 1s, Zn 2p and Cu 2p core levels were acquired. O1s and C1s spectra were acquired with a pass energy of 26 eV and an energy step size of 0.025 eV, whereas the Cu 2p and Zn 2p spectra were acquired with a pass energy of 55 eV and step size of 0.05 eV. To probe subsurface composition, depth profiling was performed by sputtering the sample stepwise with a 2 mm x 2 mm, 2 keV $Ar^+$ ion beam for a cumulative duration of 1158 s.

Initial measurements at the surface and then etching for 78 s, then six intervals of etching for 180 s taking depth profiles 2 mm x 2 mm using a 2 keV $Ar^+$ ion beam. This removed almost all oxygen signal, except for 500 °C and a small amount left at 600 °C. The Cu:Zn ratio was extracted and plotted, adjusted to one hundred percent as appropriate.

### Electron Microscopy and Spectroscopy

Transmission Kikuchi Diffraction (TKD) was performed using an EDAX Clarity Direct Electron Detector (EDAX LLC, Pleasanton, CA, USA) attached to a dual-beam Thermo Fisher (TF) Helios G5 Ga Focused Ion Beam (FIB) / Scanning Electron Microscope (SEM) (Thermo Fisher Scientific, Eindhoven, Netherlands) and analyzed with EDAX OIM 9.0 Electron Backscatter Diffraction (EBSD) software (9.1.0.85 or later, beta versions).

Some Energy Dispersive X-ray Spectroscopy (EDX) data was acquired via an EDAX ELECT PLUS EDX detector mounted on either a Thermo-Fisher Helios G3 $Xe^+$ plasma FIB (PFIB) /SEM or a ThermoFisher Helios 600i $Ga^+$ FIB/SEM (30 $mm^2$ and 10 $mm^2$ detector size respectively) and analysed with EDAX TEAM 8.0 software. Some other EDX data was acquired on a Zeiss Sigma field emission gun (FEG)-electron microscope (Carl Zeiss Microscopy GmbH, Oberkochen, Germany) with an EDAX EDX detector and analyzed using EDAX APEX software. Overall EDX compositions provided herein include C and O.

Note that for the Cu-Zn system, at.% and wt.% are very close, with an error is 0.7% at for the 50:50 alloy and 0.14% for 95:5 alloy, so can be neglected in the range of composition we are working in.

Transmission Electron Microscopy data was taken using either a JEOL 2100 Transmission Electron Microscope (TEM) or JEOL 2200 TEM (JEOL Corporation, Tokyo, Japan), both operating at 200kV equipped with JEOL EDX detectors and dark-field (DF) and bright field (BF) scanning transmission electron detector (STEM) detectors. 4D-STEM (four-dimensional scanning transmission electron microscopy) and orientation mapping was preformed using a Nanomegas ASTAR system (Nanomegas SRL, Brussels, Belgium).

### Battery Fabrication and Testing

The electrochemical properties of 10 mm diameter nanoporous Cu (np-Cu) discs dealloyed at different temperatures were tested in coin cells with metallic Li as the anode and Celgard 2500 as separator. The working electrodes tested were: Cu foil, α-brass (CuZn37), and VPD brass at 500, 600, 700, and 800 °C and the electrolyte was 80 μL 1M $LiPF_6$, EC/DMC (1:1, v/v). The coin cells were assembled in an Ar-filled glovebox. For the laboratory electrochemical testing, three replicates each for the polished flat brass, 500 °C, and 800 °C, were tested. Cycling conditions were 0.5 mA/$cm^2$, 1 mAh/$cm^2$. The cells were galvanostatically charged and discharged using a Neware battery cycler at room temperature.

### Simulation Methods

All simulations were performed with Thermocalc software version 2023a (Thermo-Calc Software AB, Solna, Sweden).

### Atom Probe Specimen Preparation and Atom Probe Tomography (APT) Measurements

Specimens for APT of np-Cu were prepared using either the Thermo-Fisher PFIB or Helios 5 CX Ga FIB described using the protocols generally in [83] and specifically in [82,84]. Where necessary, np-Cu samples were plated with Co to fill pores to prevent FIB damage, as described in [85]. When pores are not filled, FIB-induced damage both forms nanocrystalline grains around the pore edge, as confirmed by TEM, and the bottom of the pores are destroyed. APT specimens of Li-containing materials were prepared using the protocols described in [84,86].

Briefly, for cycled battery samples which were Li-plated, samples were loaded onto APT carrier puck (Cameca) in a nitrogen-filled glovebox (Sylatech GB-1200E, Sylatech GmbH, Walzbachtal, Germany) with less than 1 ppm oxygen and a dew point below -95 °C, plunged into liquid nitrogen (LN2), transferred to a pre-cooled Ferrovac ultra-high vacuum (UHV) cryo-transfer suitcase (UHVCTS) (base temperature approximately -180 °C), hereafter "suitcase" (Ferrovac AG, Zürich, Switzerland). The samples were transferred using the Ferrovac suitcase to either the FEI Helios PFIB or Helios 5 CX, using a Thermo-Fisher Aquilos-2 like cryo-stage. The cryo-stage was maintained at a temperature of -190 °C with a nitrogen flow rate of 190 mg/s. The Helios 5 CX was equipped with a cryo-manipulator (EZLift). Samples were further prepared as described above in [84], then transferred out using the Ferrovac suitcase to the APT system.

APT analysis was done using a Cameca Instruments Local Electrode Atom Probe (LEAP) 5000 XS straight-flight path instrument (Cameca Instruments, Madison, WI, USA). Data Analysis was done using Atom Probe Suite (APS) software, version 6.3.0.90 or later (Cameca). The specimen stage temperature was set to 60 K unless specified otherwise.

The run conditions for the APT data in **Figure 1(d)** were slightly varied: 60 K, 0.5 – 0.6 % detection rate (DR), 40 - 60 pJ laser pulse energy (LPE), 125 khZ 200 kHz acquisition rate, depending on specific specimen performance. The data acquisition conditions for the APT in **Figure 2(c)** and **Figure 2(j)** were similar except laser pulse energy was 40pJ. The data acquisition parameters for the three datasets in **Figure 3** were comparable.

The APT temperature datasets show that the area analyzed is a single grain, as both show a vertical crystallographic pole zone axis, as defined in [87]. The datasets were truncated to the same length and the top 2 nm removed to avoid typical fluctuations as the APT specimen evaporation stabilized.

## Electrochemical Active Surface Area (ECSA) Measurements

Samples were placed in a custom electrochemical cell and the ECSA was measured following the procedure in [88,89]. Briefly, cyclic voltammetry (C-V) scans were taken at 10 to 50 mV s$^{-1}$ in 10 mV s$^{-1}$ intervals in 0.01 M $NaHCO_3$ $CO_2$-saturated electrolyte using a three-electrode test setup, with the test sample as working electrode (WE), a Pt wire as the counter electrode (CE), and the reference electrode (RE) was Ag/AgCl [90,91]. Further information is provided in the Supplemental Information in **Figures S1 – S4**, which includes the ESCA measurement techniques.

## Nano-Computed Tomography (Nano-CT)

Specimens for nano-CT, e.g. approximately 25-30 μm diameter cylinders, were prepared using a PFIB and welded onto Cu pins having an approximately 50 μm end diameter using deposition from the platinum gas injection system (GIS) in the PFIB. Nano-computed tomography (Nano-CT) data was taken at nano-tomography beamline ID16B at the European Synchrotron Research Facility (ESRF) in Grenoble, France. The beam energy was 30 keV; the working distance was 1200 mm. Measurements were taken using a two-dimensional detector with a spatial resolution of 25 nm/pixel (actual voxel size). During individual measurement, absorption contrast micrograph was acquired every 0.1 deg. rotation step with a rotation speed of 2.5 deg./s for all 360 deg. azimuthal angles. A PyHST2 software available at ESRF was utilized for the tomographic reconstruction, initial volumetric rendering was carried out using an Avizo 9.0 software. Protocols and data processing information are provided in the **Supplementary Information**. The data was pre-processed and centered according to the standard ESRF protocols. For the remaining slices, segmentation was done using iLastik software, version 1.4.0 post1, machine learning (ML) software and subsequently processed using Dragonfly 3D World, version 2024.1. Tortuosity and pore volume fractions were calculated using that software.

# Results

## Structural and chemical characterization

An array of techniques was used to systematically examine the evolution of VPD-induced microstructural change over a 500 – 800 °C temperature range, as schematically illustrated in **Figure 1(a)**. Hereafter, the VPD samples are going to be referred to as B500 for instance for dealloying at 500 °C. X-Ray Diffraction (XRD) and energy dispersive X-ray spectroscopy (EDX) were used to provide structure and composition on bulk scales, surface composition was measured with X-ray photoelectron spectroscopy (XPS), and for analysis down to atomic scales, we used scanning and (scanning) transmission electron microscopy (SEM/(S)TEM) and associated diffraction techniques, along with atom probe tomography (APT).

Initially, the brass is a homogeneous solid solution; however, with increasing temperature, the lattice parameter of the Cu phase decreases progressively as shown in **Figure 1(b)** and **Figure S5**. This reduction is indicative of Zn depletion. For B500, a wide variety of Cu lattice spacings appeared around the (220) reflections, as shown in **Figure S5(a),** indicating lattice distortions and marked strain from inhomogeneous Zn distribution. The B500 sample showed substantial ZnO peaks compared to all other specimens, **Figure S5(c)**, supported by 2D XRD plots in **Figure S6.** To ascertain bulk Zn compositional changes, the VPD specimens prepared at each temperature were cross-sectionally polished to remove approx. 1 mm from the edge as schematically illustrated in **Figure 1(a)** and measured using EDX**.** Top and bottom surface porosity and embedding effects meant that the outer few microns could not be accurately measured with EDX because of topographic differences, so the curves are not entirely symmetrical and were trimmed to the last reliable data point on both sides and scaled to the same length. The Zn concentration near the surfaces decreased in all cases, since Zn diffuses from the open surfaces in vacuum, as shown in **Figure 1(c)**. The overall Zn concentration in the bulk decreased with higher temperatures, from an average of 37 at.% in the central region of C500 to 25 % at. in the central region of B800. Only data for B500, B700, and B800 samples are shown for clarity, but a cross-sectional graph with all curves can be found in **Figure S7**.

Surface composition by XPS and APT are plotted **Figure 1(d)** for the range of VPD temperatures. With a 100 μm diameter XPS measurement spot size, topography created by the pores and the overall surface combined to yield average composition, which gave overall higher Zn levels. APT specimens were prepared from non-porous surface regions. Both techniques confirmed the decreasing Zn surface concentrations with increasing VPD temperature, with APT showing lower Zn concentrations since the sampled volume did not include the Zn-enriched interiors of the surface pores. Additionally, XPS data shows the initial formation of significantly more ZnO for B500 versus all other samples before plating. Specifically, before plating the same XPS etching time reached a point where only Cu and Zn were present in all cases except B500, which still had measurable O, as shown in **Figure S9(c)**.

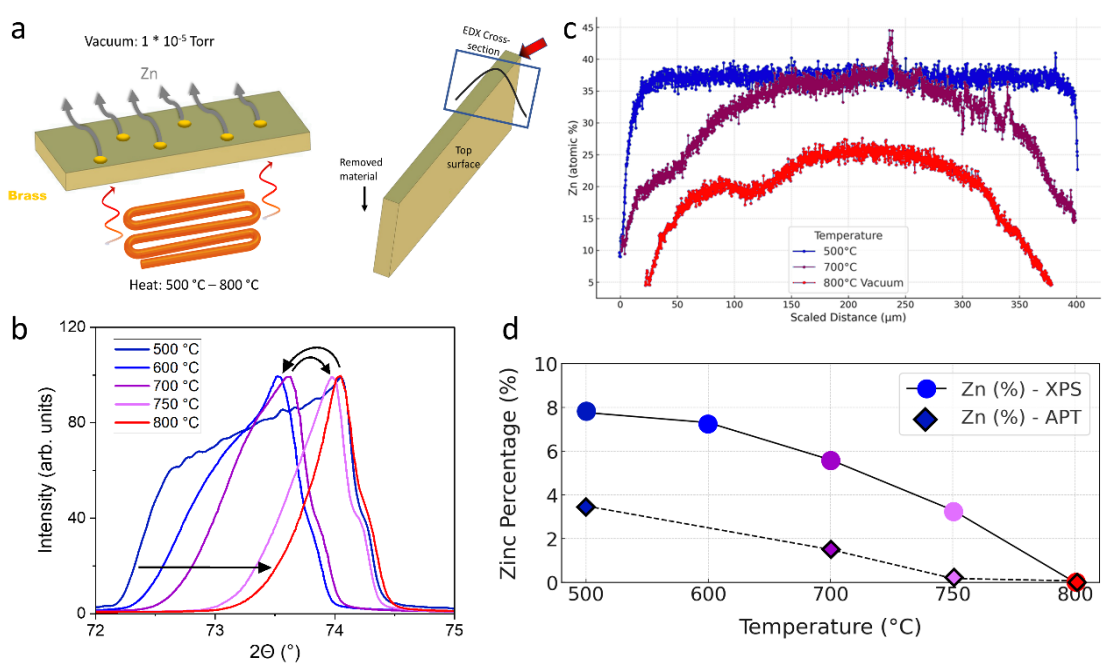

Figure 1. (a) schematic illustration of VPD-brass surface and cross-section location. (b) (220) peak of α-brass, dark arrow illustrating the shift in peak shape versus temperature. (c) 1-D EDX concentration profiles across the 0.4 mm thick cross-section, dezincified at 500, 700 and 800 ˚C. The thickness was not precisely 400 μm for each of the samples, so the width of the profile was scaled. (d) Dealloyed brass surface zinc composition as measured by XPS and APT (at.%); note that the line is just a guide to the eye.

For B500, **Figure 2(a)** is a SEM secondary electron (SE) micrograph of the highly porous surface, with relatively homogenous pore size between 1 – 3 μm. EDX mapping in **Figure 2(b)**, which shows a 11 at.% Zn overall concentration, evidences that the inner surface of the pores are enriched in Zn compared to the Cu-rich matrix. The pores also have higher Zn gradients, switching from Zn-enriched pore interiors to Cu in the higher topographic regions, with a small APT dataset illustrating this in **Figure S8(a)** and **Figure S8(b)**. An approximately 20% O at.% composition in the Zn-rich areas was present, indicating possible oxidation during vacuum dealloying. Diffusion or trapping of Li within ZnO could modify the battery performance [81,92]. Comparative surface O and Zn distributions for B500 surface are shown in **Figure S9(a)** and **Figure S9(b)** for comparison. A top-down APT analysis, **Figure 2(c),** reveals a Zn concentration gradient from approx. 3.5 at.% to 4.5 at.% from the near surface down to 150 nm below the surface**.**

For XPS measurement, all dealloyed samples were briefly sputtered with $Ar^+$ ions to remove residual C and O until the C signal fell to near background. However, for the B500 XPS surface composition measurement, a substantial O percentage remained present even after a long sputtering run when the C signal declined to near background in **Figure S9(c)**, indicating a thick residual surface oxide, as substantiated in later APT from a surface pore boundary in B500, as shown previously in **Figure S8(a-b).**

Transmission Kikuchi diffraction (TKD) was performed in the SEM at cryogenic temperature (-190 ˚C) on a lamella lifted out from the surface also at cryogenic temperature as this is the workflow required later in lithiated condition (see Methods). Since the pores were expected to contain a very thin Zn layer, cryogenic preparation was used to minimize FIB damage to the bottom pore walls. **Figure 2(d)** reveals several individual grains, numbered 1 – 5, along with shallow pores on the surface visible under a Pt cap layer. Larger pores imaged below the surface have further coalesced and coarsened during the VPD process. Zn is found on the inner surface of the closed subsurface pores, as confirmed via STEM-EDX maps in **Figure 2(e)**. Hollow pores inevitably suffer from focused ion beam (FIB)-induced degradation during specimen preparation and etching of the bottom of those pores, evidenced by the streaking shown in **Figure 2(d)** and **Figure 2(e)** beneath those pores, explaining why Zn is present only at the top. The Zn inside the pores can be explained by the condensation of evaporated Zn still within the pores as the specimen cools down. However, there was no O present in a deeper pore, see **Figure S9(d)** showing the Zn distribution to be homogenous around the top of the pore, where undamaged by the FIB, and minimal O distribution throughout the matrix and pore edge in **Figure S9(e)**. This supports that the O in the open pores is from the residual gases in the vacuum furnace.

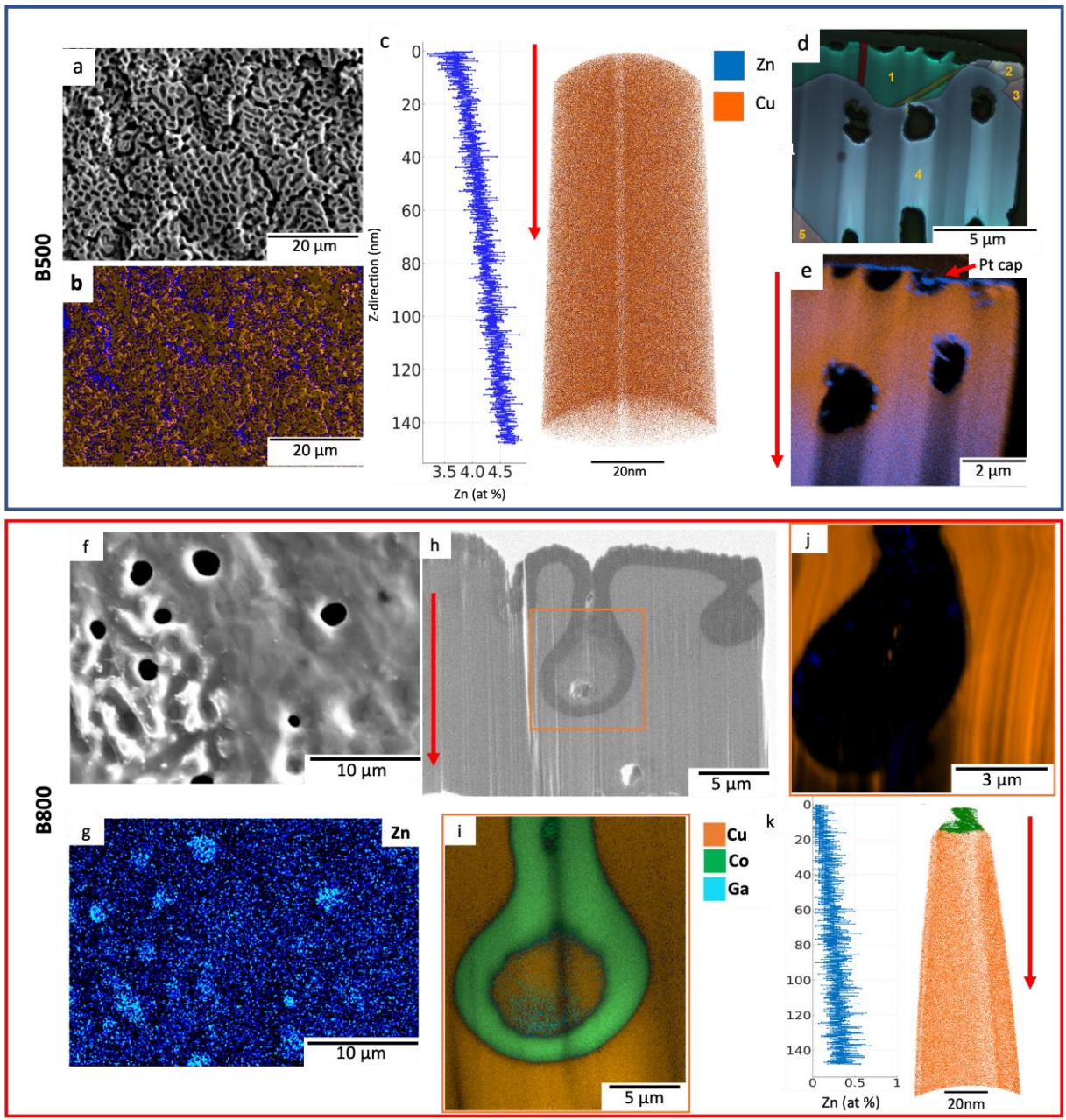

**Figure 2.** Images of the surface of dezincified brass sample B500, prepared at 500 °C (a) top surface SEM image (b) EDX overlay for 500 °C of Cu and Zn relative abundance (average surface composition, multiple scans, atomic %: 88% Cu, 6% O, 5% Zn)  (c) Quantitative APT 1-D Z-axis elemental profile showing Zn gradient down from surface; vertical line is zone axis indicating single grain (d) TKD grain map overlaid on SEM image, showing 5 grains in this lamella, along with inset Zn TKD pattern showing Zn inside closed pores, along with FIB damage at the bottom of pores (e) TEM EDX elemental map of lamella, illustrating Zn inside closed pore; dezincified brass sample B800, prepared at 800 °C (f) top surface SEM view with lower visible pore density (g) lamella prepared from Co-plated surface to fill pores, Co is darker (h) TKD image of R1 showing that the area is a single grain (i) EDX Zn distribution (87% Cu, 11% C, 2% O, 0.6% Zn, atomic%) top surface, concentrated in pores (j) quantitative 1-D Z-axis elemental profile showing Zn gradient down from surface, denoted with Co cap, with vertical line is zone axis indicating single grain (k) SEM EDX view, Co-filled open pore in R1, illustrating lack of Zn.

Regarding B800, the SEM surface image in **Figure 2(f)** shows a heterogenous pore size distribution with a relatively smaller fraction of surface openings. This is consistent with the pore characterisation by nano-CT detailed in **Figure S10(a-f)**. To summarize, with increasing temperature, pores under the surface tend to be more unconnected, i.e. decreasing tortuosity, and larger pores with increasing volume. This trend is consistent with the electrochemical active surface area (ESCA) as in **Figure S1** and supporting **Figures S2 – S4**. B800 had higher ECSA, indicating that it has more disconnected, larger pores and lower tortuosity than B500, matching the nano-CT trends. The surface Zn concentration was approximately 0.6 at.% according to EDX mapping, **Figure 2(g)**, with Zn primarily concentrated on the surface of the open pores.

The B800 sample was subsequently electroplated with Co to avoid FIB damage to inner regions of the larger open pores during specimen fabrication [77,85]. **Figure 2(h)** is a SEM side view image of a lamella prepared from that sample with visible filled Co-filled pores, with a highlighted region in the orange box containing one filled surface pore. **Figure 2(i)** display**s** a SEM EDX map of the boxed region, highlighting the bulk area, mapping the Cu distribution with negligible Zn around the Co-filled pore. The Zn could have been depleted during electroplating or otherwise masked by inherent roughness and faceting inside the pore. Note that the Cu inside the Co was redeposited during the FIB-lamella preparation process. **Figure 2(j),** a TKD map of the boxed region, demonstrates that the matrix around the pore is a single grain without nanocrystalline FIB-induced damage at 40 nm TKD resolution. APT analysis in **Figure 2(k)** reveals near complete Zn removal at the surface, delineated by the Co cap, with an increasing Zn gradient downward into the bulk**.** Additional APT analyses at increasing distance from a subsurface pore (≈2 μm) are summarized in **Figure 3**. At the edge of the pore, the Zn composition was approximately 5 at.%, whereas even 500 nm – 1 μm away, the Zn content is close to the bulk composition observed by APT and XPS, which reverts to the bulk concentration farther from the pore.

This gradient clearly establishes higher Zn levels on the open interiors of surface pores exposed for Li plating.

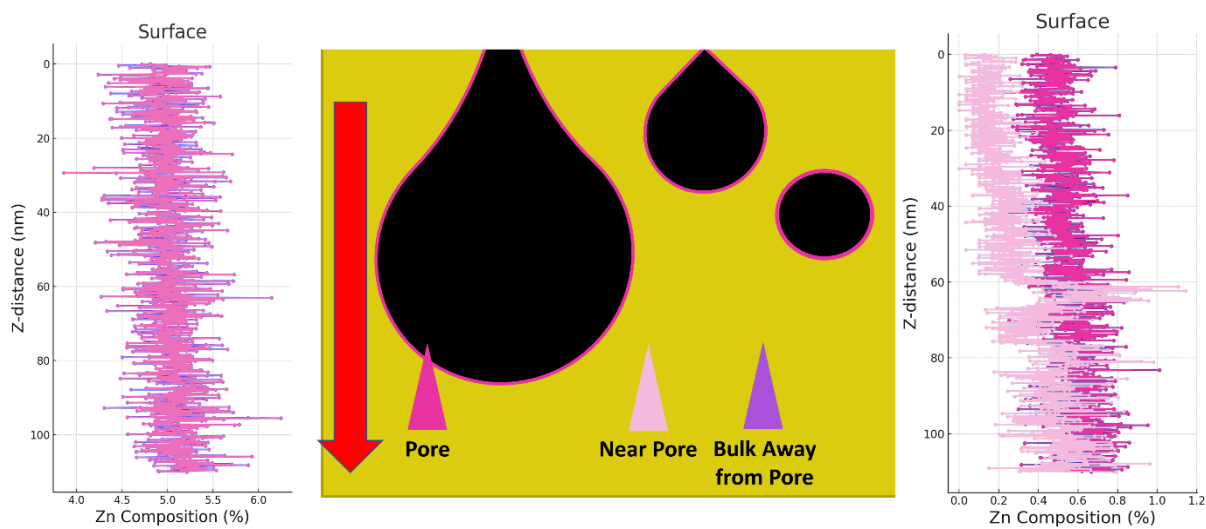

**Figure 3.** APT composition gradients at various points on the edge of a pore (left) and nearby with respect to pore edge (right, less than 1μm), and far from (right, greater than 1 μm) a pore in a 750 °C dealloyed brass lamella. Inset diagram shows relative locations of APT specimens, highlighting increased Zn concentration in open surface pores

## Battery Testing and Performance

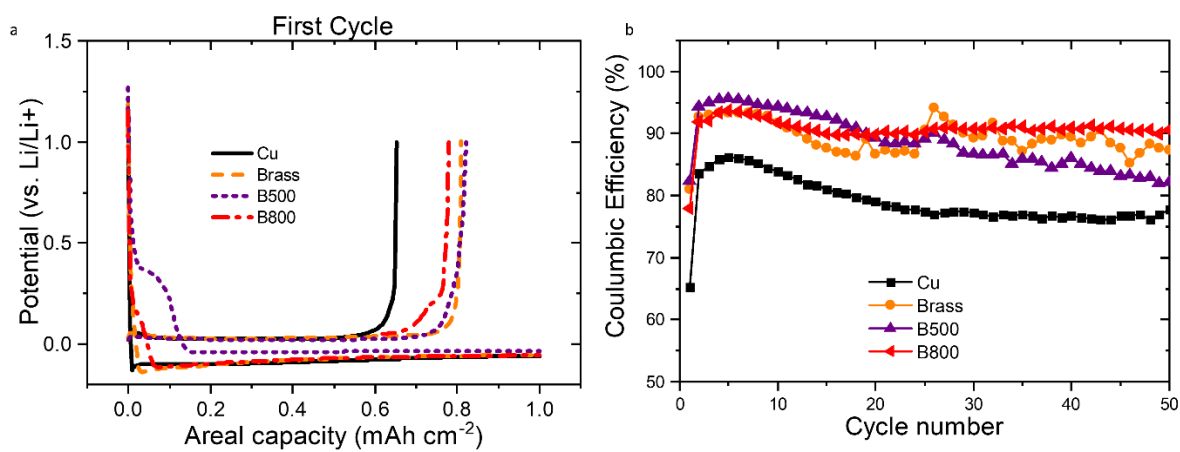

**Figure 4.** Battery performance data (a) comparison of first-cycle performance of different current collectors / working electrodes, illustrating in the left lower corner that the higher the Zn content of an alloy, the more Li-Zn alloying occurs during the first cycle (b) Comparison of overall CE for 50 cycles for different current collectors: Cu, flat polished brass, B500, and B800

Li metal half cells were prepared with 10 mm discs of brass subjected to VPD at 500 – 800 °C. Cu foil was used as a state-of-the-art current collector reference sample, along with a non-dealloyed flat polished brass piece. **Figure 4(a)** plots the potential vs. Li/Li$^+$ as a function of the areal capacity. The sharp dip below 0 V represents the overpotential – and corresponding thermodynamic driving force – required to initiate Li nucleation and plating on the pristine Cu surface. There is a kink in the lower left corner of the other curves, between 0.5 and 0 V, that indicates Li-Zn alloying. Stronger Li-Zn alloying reaction is observed in samples dealloyed at a relatively lower temperatures, i.e. with a higher Zn composition. Alloying is also followed by Li plating.

**Figure 4(b)** shows the evolution of the Coulomb efficiency (CE) at increasing charge / discharge cycles for the pristine Cu, flat polished brass, B500, and B800. B800 displays the lowest degradation and stable trends. B500 linearly degrades over the first 50 cycles below both flat, polished brass and B800. The flat, polished brass CE stabilizes quickly but displays instability, whereas B800 does not. The Cu substrate does not reach similar CE and degrades quickly over the first 20 cycles. Additional CE for all tested dealloying temperatures are given in **Figure S11(a)** and first – third cycle curves for B800, flat polished brass (as in [93]), and Cu foil are given in **Figure S11(b-d)**.

Electrode microstructure

Characterization of the plated flat bras, B500, and B800 after 100 cycles prepared by cross-sectional cryo-FIB-SEM is reported in **Figure 5**. Cryogenic conditions are necessary to prevent possible Li migration during electron imaging; Li appears dark in the SEM images versus Cu because it has low secondary electron yield [94–96]. Li plates relatively uniformly on the flat polished brass as seen in **Figure 5(a),** while forming an intermetallic containing the ternary Laves phase $Cu_3ZnLi_2$ [93]. The enlarged view on the brass-Li interface in **Figure 5(b)** reveals a thin layer with an intermediate contrast corresponding to the intermetallic layer containing the $Cu_3ZnLi_2$ Laves phase [93], which formed as an intermetallic. The B500 plated unevenly, as shown in **Figure 5(c)**, even when most of the Li peeled off when the cells were opened. **Figure 5(d)** is a close-up on a Li-filled pore, confirming that Li penetrated sub-surface. On B800, Li plated uniformly and filled the open surface pores as shown in **Figure 5(e)** and the close-up in **Figure 5(f)**. The dark contrast suggests that some sub-surface pores are also filled with Li, likely when interconnected with the surface.

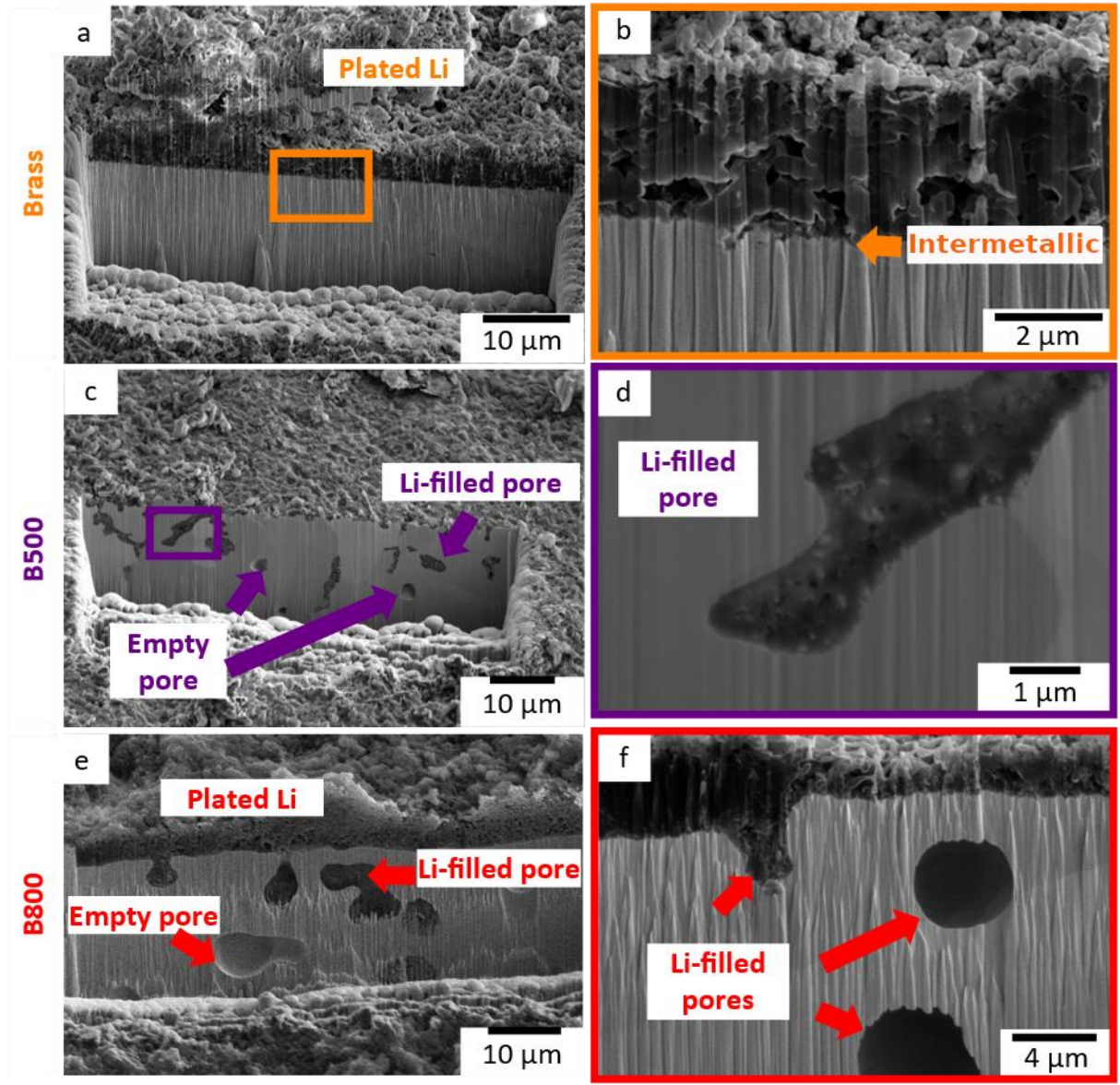

**Figure 5.** Secondary electron imaging on cryo-FIB cross-sections of Li-plated after 100 cycles on (a-b) flat brass; (c-d) B500; (e-f) B800.

Electron-transparent lamellae were FIB-prepared to perform TKD on these three samples. The metallic regions were the analytic focus; therefore, these lamellas were milled to slightly less than 200 nm final thickness, precluding Li indexing, which typically requires 1.2 – 1.5 µm Li thickness. In **Figure 6(a)**, the Li plated uniformly on the polished brass with a deformed nanocrystalline grain layer, which subsequently formed an intermetallic [93]. For each TKD dataset, after spherical reindexing, all points indexed as the matrix with no second phase observed, except the deformed intermetallic layer, as explained below. As such, another metric, the grain reference orientation deviation (GROD) was computed. GROD quantifies the angular deviation of each point from the average orientation of the nearest grain, with higher values indicating high stored energy and deformation-induced strain gradients. This value helps distinguish between the original microstructure and potentially lithiated or deformed areas. GROD analysis in **Figure 6(b)** highlights a 200-nm-thick deformed layer, which contains nanoscale grains indexed as the same phase as the substrate. However, the deformation and nanoscale grain size precluded complete indexing; even after optimized spherical reindexing, only small parts of this deformed layer had confidence index (CI) > 0.15, a threshold chosen to exclude spurious points. The substrate immediately beneath this layer exhibits higher GROD values consistent with a strain gradient, potentially induced by compositional changes, e.g. Zn depletion below the layer and Li-Cu-Zn alloying [97,98].

For B500 in **Figure 6(c)**, the Li appears porous and did not plate uniformly. The GROD orientation map in **Figure 6(d)** reveals three large grains with no gradients or changes in either the surface or the open pores indicative of Li-Zn alloy layers or stress gradients; however, one grain boundary (GB), marked by an arrow, highlights Li penetration along it. Preferential lithiation down GBs was previously reported and could contribute to damaging the current collector from localized strain accumulation [77]. The B800 lamella, **Figure 6(e)**, contains one empty pore on the left and two filled ones on the right and TKD confirms that between pores, the current collector is a Cu single grain. A GROD image near the empty pore is displayed in **Figure 6(f)**; at the top, there is no evidence, e.g. changes in values, indicating possible strain gradients or thin Li-Zn layers around the pore walls. The GROD values inside and outside the pore in the top half remain similar, indicating an absence of Li-Zn alloying or strain gradients. As previously demonstrated in **Figure 2(e),** when preparing a lamella by FIB milling, residual Zn, if present, is typically preserved only around the upper half of the pore walls. In contrast, the significant, seemingly random GROD deviations in the bottom half of the pore are FIB-induced artifacts, which typically occurs when milling empty pores. Similarly, for the Li-filled pore in **Figure 6(g),** which shares the GROD scale with **Figure 6(f)**, no thin layer similar to the deformed layer in **Figure 6(b)** appears at the pore boundary between the Li and the B800 matrix.

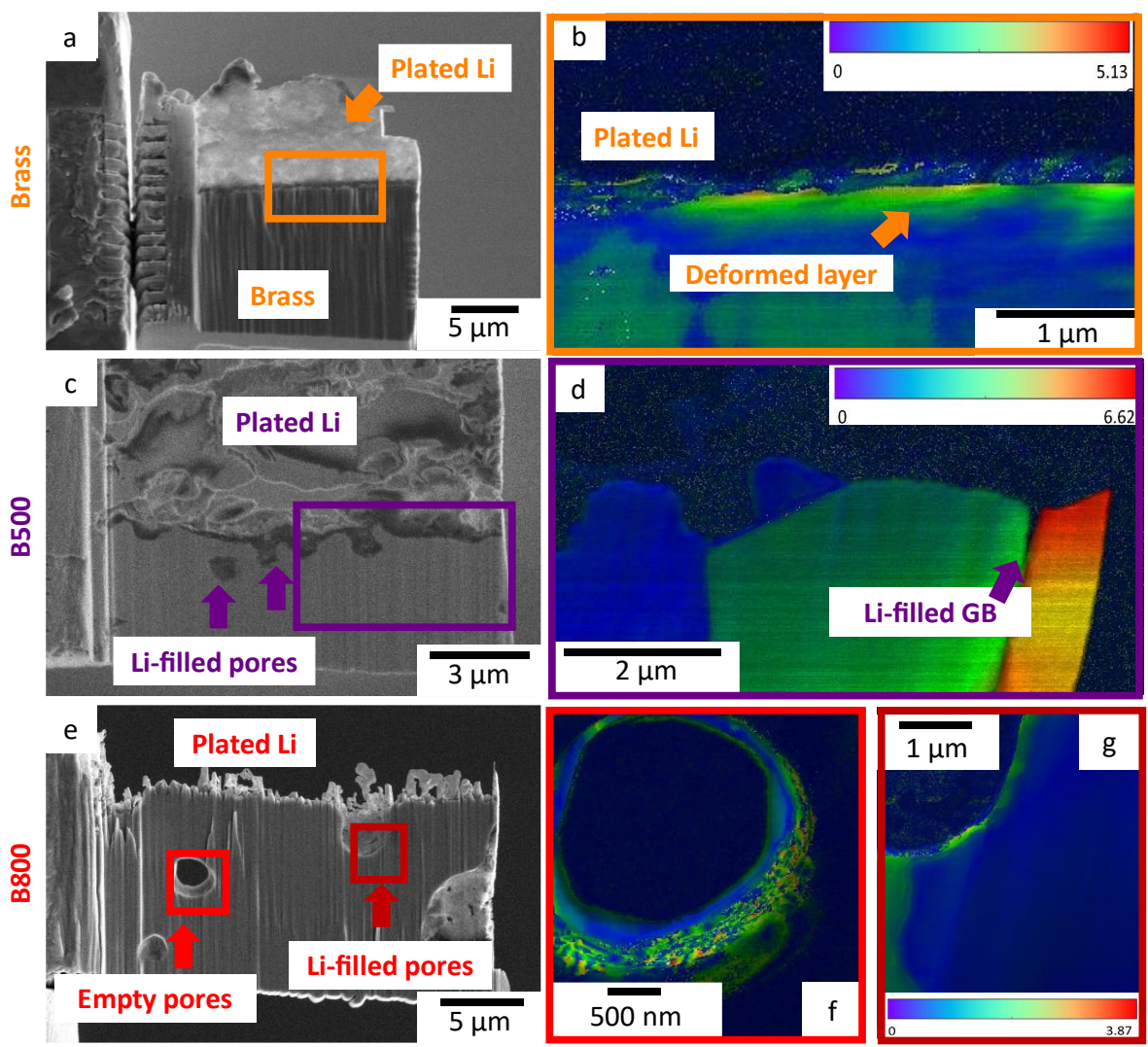

**Figure 6.** SEM cross-sectional views and TKD of Li-plated lamellae prepared from samples in **Figure 5**. (a) Flat polished brass lamella, showing the deformed layer (b) GROD map of flat polished brass layer showing thin deformed layer (c) B500 lamella showing inconsistent Li-plating (d) GROD map of B500 showing grains and Li-filled GB (e) B800 lamella with thin remaining Li and empty versus Li-filled pores (e) B800 lamella with empty and filled pores (f) GROD map of B800 empty pore with bottom, scale bar in (g) GROD map of B800 Li-filled pore

XPS depth profiles of the stripped B500 and B800 samples were performed. Initially, before cycling, XPS measurements showed B500 had approximately 8 % at. Zn at the surface, while B800 had approximately 1 – 2 % at. Zn overall, but APT showed < 1% in surface areas not pores. The XPS results are detailed in **Figure S12** and **Figure S13** respectively. For both, the percentage of Cu and Zn versus Li did rise as the depth increased relatively in proportion to its presence in the substrate, indicating that excess Zn did not accumulate in the plated Li. Note that the Li percentage remained above 70%, and therefore did not mostly reach the unreacted substrate, showing incomplete Li electrochemical stripping. Since the Cu and Zn percentages increased in relative constant ratios to their presence in the substrate, likely the metal came from physically higher areas of the substrate exposed to the ion beam. As the surface topography is more pronounced in B500 with more pores, as shown in **Figure 2(a) and Figure 2(f),** the XPS data is consistent with this hypothesis.

XPS assisted in deconvoluting the species involved in the electrochemical process and the reaction products. Significant content of F and P were found in the Li, derived from the LiPF$_6$ electrolyte, where literature shows that LiPF$_6$ decomposes to LiF, HF and PF$_x$, providing free F [99]. For B500 in **Figure S12**, as the etched depth increased, the Cu bonding state did not change; however, the Zn bonding states changed with depth, corresponding to Zn halides, Zn phosphides, and Zn$^{2+}$ consistent with ZnO [100]. This bonding structure is consistent with electrolyte decomposition and Zn-P and Zn-F bond formations, along with Zn-O in the B500 case [101]. Cu is more noble and so did not form significant bonds with either F or P. For B800 in **Figure S13**, the Cu remained in a dominant Cu$^{2+}$ state throughout the depth profile. The low content of Zn (< 2%) remained in a mix of Zn halide and Zn$^{2+}$, with low signal-to-background ratios making interpretation of bonding changes challenging.

# Discussion

## VPD: controlling pore size and Zn composition

The differential vapor pressure of binary alloy systems, particularly Cu-Zn, allows for easy VPD of Zn to form porous structures. CuZn37 can serve as an effective substrate to produce these. VPD is advantageous over chemical dealloying because it can result in effectively complete removal of the sacrificial metal in a binary system in a "dry" manner, e.g. without introducing any liquids, solvents, or acids. Liquid-based chemical dealloying is not as effective, particularly for lower concentrations of the sacrificial metal like CuZn37, as documented in [43]. It also results in less waste, as the dealloyed element can be recovered by condensing the metal from the exhaust.

There is qualitative agreement with the expected pore and ligament coarsening rate based on the theory reported in Ref. [102–104]. However quantitative analysis cannot be performed at this stage, since some data is missing for B500 and B800. Ref. [100] proposes regimes with a different dominant physical process and driving force for pore formation: (i) evaporation-condensation [60]; (ii) grain-boundary diffusion and/or a combined mechanism between grain-boundary diffusion and evaporation-condensation (rare in VPD [104]); (iii) bulk diffusion; (iv) surface diffusion [102,105,106].

As revealed by nano-CT and SEM, lower temperatures yield smaller but more connected pores with higher tortuosity and pore connectivity, while higher temperatures lead to pore coalescence, and coarsening, which can be extensive at higher temperatures. At 500 °C, the numerous, small and shallow surface pores are Zn-enriched, indicating that surface diffusion (iv) dominates in this temperature range. This is supported by APT showing that Cu-based regions have low Zn, and it increases slowly moving away from the surface. Zn is also concentrated on the inner wall of Kirkendall pores visible in **Figure 2(d)** and **Figure 2(e)**, so there is slower bulk diffusion happening as well. As the temperature increases towards 800 °C, the pore structure evolves to larger, deeper, less connected pores with increased volume and lower tortuosity. That suggests a shift towards a mixed transport mechanism for Zn removal, while the significant reduction in Zn in the bulk indicates that enhanced bulk diffusion (iii) becomes more significant. Additionally, at elevated temperatures, the high vacuum increases efficient evaporation-condensation processes (i) inside the deeper pores and results in full conversion of bulk brass to Cu at 900 °C (see **Figure S20).**

To better understand the influence of the temperature on the kinetics of VPD in the Cu-Zn system, Thermo-Calc and DICTRA simulations were employed to model the evolution of composition profiles under various temperature and pressure conditions and provide insights into the fundamental mechanisms governing the dealloying process and the thermodynamic driving forces at work. These are detailed in Suppl. Information **Figures S14** - **S19**. However, at this stage, discrepancies between the performed simulations and the experimental data at high temperature, particularly **Figure S16(b)**, preclude using these thermodynamic simulations in a predictable way. Yet, our work provides the basis for choosing the appropriate temperature to create a specific type of porous structure with a selected Zn content. Since Zn content emerged as the single most relevant variable for battery performance in the porous Cu-Zn system, temporal evolution of pore structures would provide secondary tuning mechanisms to further improve performance. However, factorial experimental design of temperature versus time conditions were beyond the scope of this study.

### Battery Performance

The electrochemical performance data reveals interesting differences between the dealloyed materials prepared at different temperatures. The B800 demonstrates superior cycling performance compared with both the reference sample and the sample dealloyed at lower temperatures, despite having fewer visible surface pores. We attribute this enhanced performance to several factors: first, the deeper pores are volumetrically larger and completely filled with Li, as shown in **Figure 5(e), Figure 5(f),** and **Figure 6(e).** Although the deeper pores in B500 do fill with Li in **Figure 5(c)**, the surface plates non-uniformly as in **Figure 5(c-d)** and **Figure 6(c).** B500 had significant O observed in the surface layers, with XRD showing significant ZnO peaks in **Figure S5(c)**. The B500 sample showed substantial ZnO peaks versus all other specimens with XPS in **Figure S9(c),** additionally demonstrated by the O signal in APT (**Figure S8**) and XPS (**Figure S9(c)**). The B500 surface EDX map showed 11 at.% O and 12 at.% Zn with correlated distributions as seen in **Figure S9(a-b)**, which was not seen in the deeper, closed pores in **Figure 2(e)** and **Figure S9(d-e)**. Li diffusion in Zn is known to be close to ca. $2 \times 10^{-10}$ cm$^2$/s at 25 °C [107] whereas for the case of ZnO it is close to ca. $5 \times 10^{-19}$ cm$^2$/s at the same temperature [108].

Considering this fact, while B500 is more tortuous, Li-migration and plating is inherently slowed by the initial ZnO barrier. The lower out-diffusion of Zn at 500 °C is controlled by a different mass transport regime, which allows formation of surface ZnO even under vacuum during furnace cooling.

The first cycle charge-discharge curves in **Figure 4** clearly demonstrate a trend for more Li-Zn alloying with decreasing dealloying temperature, with the 500 °C dealloyed brass showing the highest Li-Zn alloying, with supplemental data in **Figure S11**. That trend, along with worsening electrochemical performance over time for the lower temperature dealloyed samples indicates that excessive Zn becomes detrimental for performance. APT and other studies confirm that there is only either Cu or CuZn at the surface, e.g. 1 – 2 at.% for 750 °C, with higher residual Zn concentration in the pores, ~ 5 at.%. Since D800 with ~1 at.% Zn surface composition had the best performance, very low concentration of Zn at the surface, with higher concentrations on the inner open pore walls is helpful for ensuring improved plating and dendrite suppression, as are larger pores which are open to the surface, even if the number of openings is quantitively smaller, e.g. for the 500 °C case in **Figure 2(a)** versus the 800 °C case in **Figure 2(f).**

The comparable ESCA values between 500 °C and 800 °C indicate that surface area alone does not determine electrochemical performance. Rather the combination of pore architecture and volume, as well as the Zn concentrations and surface chemistry, appear to create more favorable conditions for Li plating in the higher temperature dealloyed materials. Overall, this data suggests that optimizing dealloying temperature allows for careful control of both alloy composition and pore network size and depth, which together influence Li plating behavior and battery capacity retention.

## Conclusion and Future Work

This work establishes VPD temperature as the critical parameter controlling both porosity and residual Zn content in Cu-Zn current collectors. By systematically dealloying α-brass (Cu63Zn37) at temperatures between 500 – 800 °C, we showed that surface Zn composition decreases from 4 -8 at. % to < 1 at. % with increasing temperature. Zn mass transport and out-diffusion at 500 °C operates in a different mass transport regime, which allows significant surface ZnO formation even in high vacuum than higher temperatures. With increasing temperature, pore structure tended to evolve from smaller, connected pores networks towards larger, less-connected pores with higher surface area. Thermodynamic modelling revealed some additional Zn transport mechanisms that became more significant at higher temperatures.

Electrochemical testing in anode-free Li-metal half cells demonstrated that minimizing Zn surface concentration to ~1% surface concentration provided optimal performance. VPD-prepared brass at 800 °C with ~ 1 % surface Zn content showed stable CE > 90% over 50 cycles, better than all other tested samples; the 500 °C showed rapid CE degradation to around 70%, which can be attributed to the thicker initial ZnO surface layer. The inverse correlation establishes an upper limit of 1-2 at.% surface concentration for optimal Li plating and performance.

Importantly, this work demonstrated that commercially available α-brass, with Zn concentration near the parting limit, can produce porous current collectors. The VPD process offers a scalable, simple, environmentally benign method to produce current collectors, avoiding the waste from wet etching.

Future work could include further characterization of the surface alloying behavior of Li-Zn within the pores of VPD-prepared brass at 500 °C and 800 °C with TEM and APT to quantify the actual Zn inside open and closed pores. Further synchrotron nano-CT studies are needed to get pore size and structure for those two temperatures, as well as in-situ nano-CT studies to observe operando VPD and temporal porosity evolution, which would provide additional possible increases in electrochemical active surface area. Even lower dealloying temperatures and longer dealloying times offer different mass transport regimes, which would provide further insight into the process kinetics. Operando synchrotron X-ray spectroscopy studies (e.g. EXAFS, XANES) to examine Zn transport during Li plating and stripping would enhance understanding of its role in ensuring Li plating uniformity, particularly during the first cycle where Li-Zn alloying is electrochemically observed.

## CRediT Author Contribution Statement

## Acknowledgments

The authors would like to thank Rainier Lück, Ralf Selbach, Tristan Wakefield, Mario Bütow, and Simon Sprengel in the MPIE mechanical workshop for help with substrate and holder fabrication. The authors would like to thank Jürgen Wichert for assistance with VPD and furnace holder development. The authors would like to thank Andreas Sturm for his assistance with the MPIE FIB facility; Uwe Tezins for assistance with the MPIE APT facility; and Heidi Bögershausen for her assistance with SEM, EDX, EBSD, and sample preparation. The authors would like to thank Benjamin Breitbach for his assistance with XRD measurements. The authors would like to thank Petra Ebbinghaus for her assistance with AFM measurements. The authors would like to thank Dr. Se-Ho Kim and Dr. Aparna Saksena for their helpful discussions regarding electrochemical dealloying and the kinetics of transport. E.V.W, Ö.Ö., and S.L.W. acknowledge the European Synchrotron Radiation Facility (ESRF) for provision of synchrotron radiation facilities under proposal number MA-6182. We also would like to thank beamline scientists Prof. Dr. Gustova Pinzon and Dr. Katrin Bugelnig for their valuable assistance and support at beamline ID-16B.

## Funding

E.V.W. and B.G. would like to thank the DFG for their funding through the Leibniz Prize. E.V.W., A.E.Z., and B.G. are grateful for funding form the ERC for the project SHINE (ERC-CoG) #771602. T.M.S. gratefully acknowledges the financial support of the Walter Benjamin Program of the German Research Foundation (DFG) (Project No. 551061178). M.J.K. thanks the Fraunhofer and Max Planck cooperation project MaRS (Critical Materials Lean Magnets by Recycling and Substitution) for their financial support. J.L. was supported by the Natural Sciences and Engineering Research Council of Canada (NSERC) Discovery Program (RGPIN-2023-03655), Canada Foundation for Innovation (CFI), BC Knowledge Development Fund (BCKDF), and the University of British Columbia (UBC). X.C. gratefully acknowledges the DFG for their support through the Collaborative Research Centre/Transregio (CRC/TRR) 270 HoMMage-Z01 project. Y.J. and M.P.S. are grateful for funding from EU Horizon Research and Innovation Actions under grant agreement number 101192848 (FULLMAP). S.L.W. thanks the financial support from the Alexander von Humboldt Stiftung through the AvH fellowship. Y.J. acknowledges funding by the European Union through EIC grant no. 101184347, Heat2Battery. B.G. and Y.K. thank the DFG for support through DIP Project No. 450800666. Ö.Ö. would like to acknowledge funding through the International Max Planck Research School for Sustainable Metallurgy (IMPRS SusMet). Y.Z. and G.D. would like to acknowledge financial support by the Max-Planck-Gesellschaft.

Open access funding enabled and organized by Projekt DEAL.

## Data Availability

The data is available from the authors upon reasonable request.

## Conflicts of Interest

The authors declare no conflicts of interest.

# Supplemental Information

## Electrochemical Chemically Surface Area (ECSA) measurements

| Sample | C (mF) | $E_@$ (V) | $R^2$ | ECSA (cm$^2$) |
|---|---|---|---|---|
| 800_Argon | 0.03399 | -0.1 | 0.971 | 1.22 |
| 800_Vacuum | 0.04979 | -0.1 | 0.993 | 1.84 |
| 750 | 0.03065 | -0.1 | 0.999 | 1.11 |
| 700 | 0.0398 | -0.1 | 0.982 | 1.42 |
| 600 | 0.06245 | -0.1 | 0.962 | 1.14 |
| 500 | 0.03165 | -0.1 | 0.987 | 2.24 |

**Figure S1.** Calculation parameters and results of ECSA for the samples

## ESCA measurement

**Estimating the electrochemical surface areas (ECSA) from cyclic voltammetry (CV) curves by double-layer capacitance measurements.**

ECSAs for the porous Cu-Zn electrodes relative to polycrystalline Cu foil were determined by measuring double layer capacitances [86,106,107]. CV curves were performed in the same electrochemical cell as in 0.01 M NaHCO$_3$ electrolyte. A standard three-compartment electrochemical cell was used: (1) the Cu based electrodes were used as the working electrodes with a 0.216 cm$^2$ geometric area exposed to the electrolyte, (2) a Pt wire separated by porous ceramics as the auxiliary electrode was used as the counter electrode and (3) an Ag/AgCl electrode (Pine) was used as the reference electrode through a Luggin capillary. The reference electrode potential is converted to the RHE reference scale using E (vs RHE) = E (vs Ag/AgCl) + 0.201 + 0.0591 V × pH.

We choose the same potential window for all samples but cannot avoided Faraday's zone for all samples completely (see **Figure S2**) due to different components. Capacitance is calculated using the surface area of the piece, either 3 x 7 x 0.4 mm or a 10 mm disc x 0.4 mm thickness. However, for rough comparison [85], double-layer charging and discharging (no Faradaic process) were assigned to the potential window of 0.5 – 0.7 V vs RHE according to CV curve of the 600 °C treated sample. The current in the potential window was plotted against the different scan rate of the CV curves (**Error! Reference source not found.**). The slope of the linear regression gives the capacitance (**Figure S4**). In combination with the empirically measured $C_{Cu}$ = 29 µF·cm$^{-2}$ for planar polycrystalline Cu [85,86,106,107], we can determine the electrochemical active surface area by $A_{ECSA}$ (cm$^2$) = $C_{DL}/C_{cu}$, where $C_{DL}$(µF·cm$^{-2}$)=I(µA·cm$^{-2}$) / $v$(mV/s), summarized in [86].

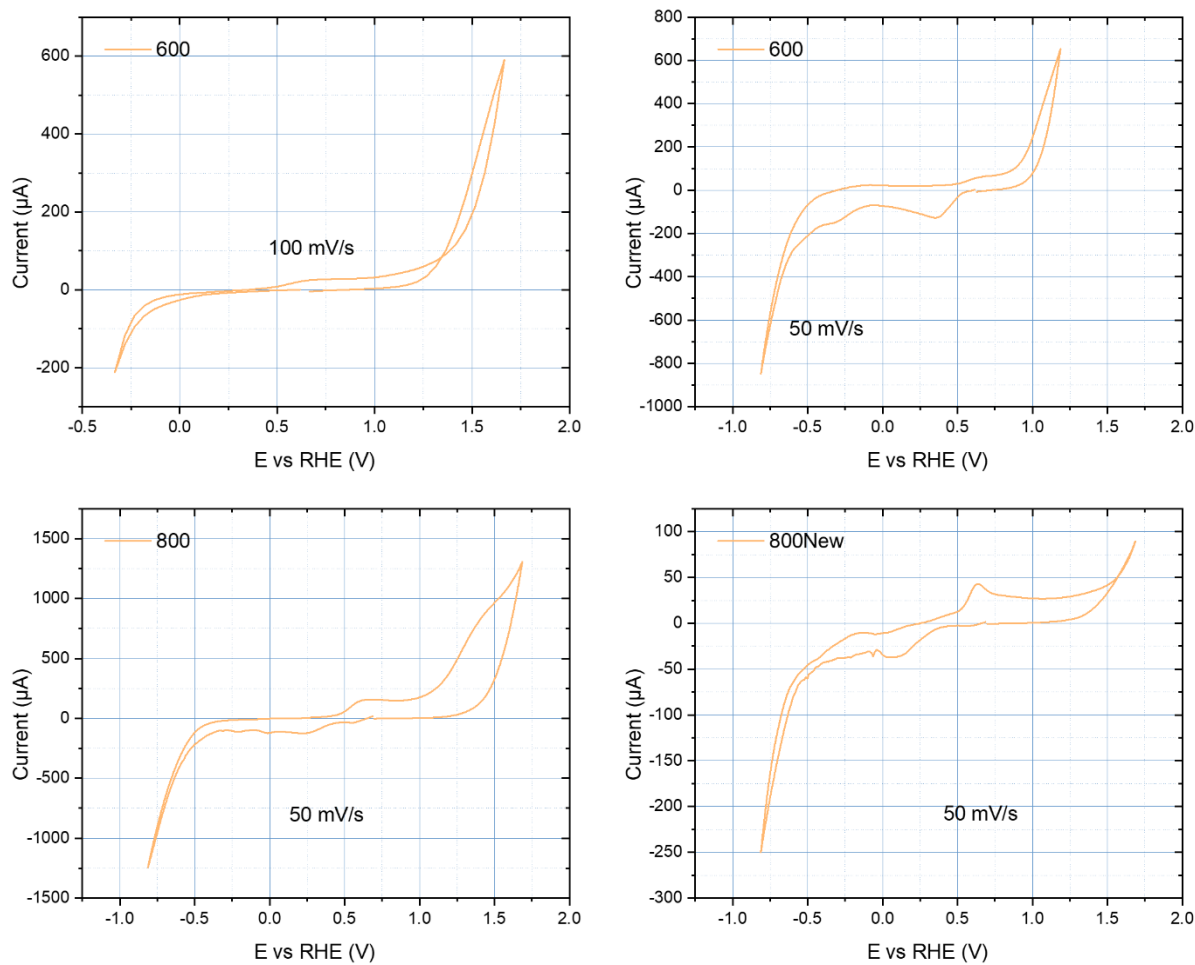

**Figure S2.** A cyclic voltammetry scan taken at 100 mV s−1 in 0.01 M NaHCO3 electrolyte.

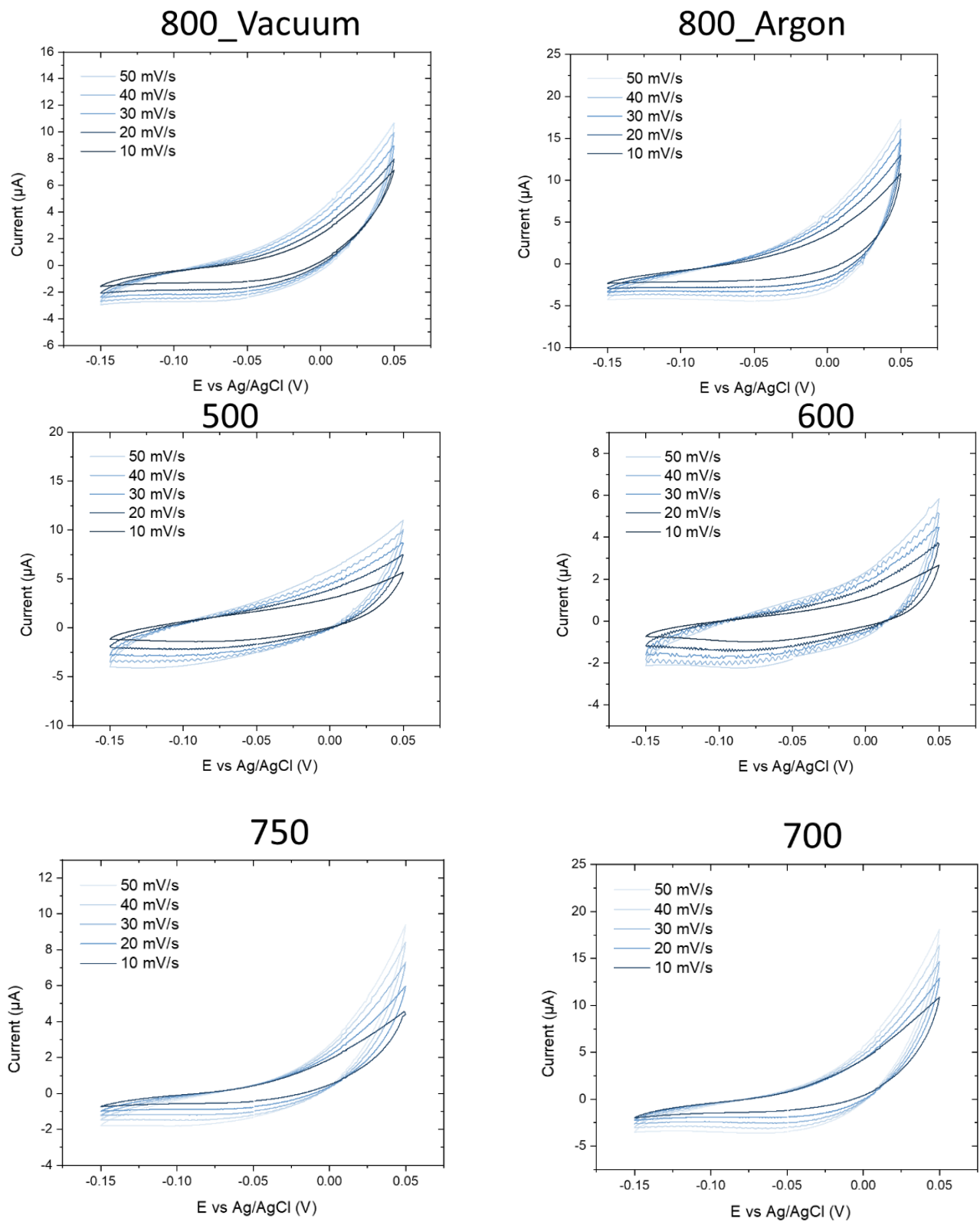

**Figure S3.** Electrochemical CV curves and various scan rates for determining the double layer capacitances of different samples. 3 electrode test setup CE: Pt wires, RE: Ag/AgCl, CO2 saturated 0.01M NaHCO3 solution, room temperature.

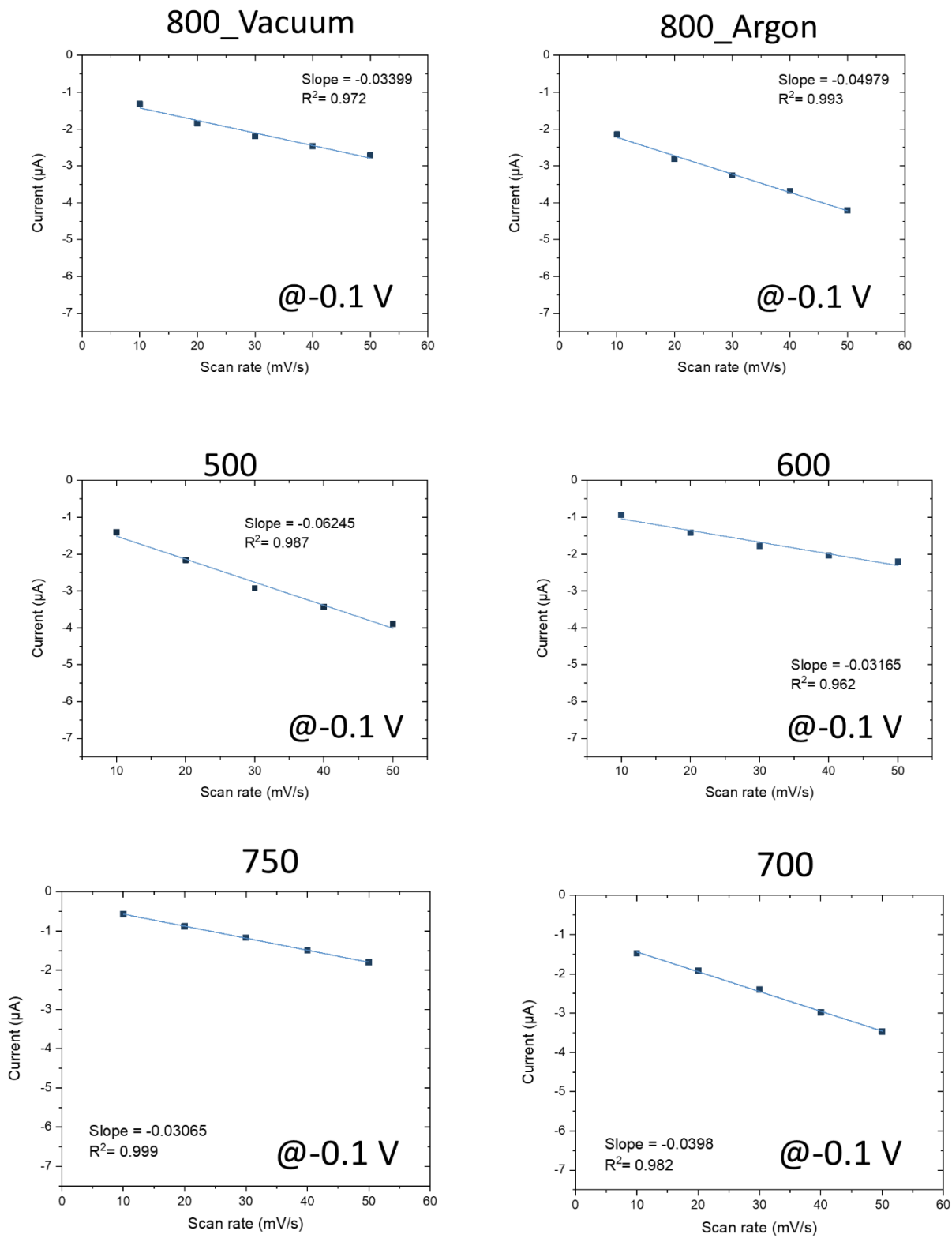

**Figure S4.** Current due to double-layer charging plotted against C-V scan rates

## X-Ray Diffraction (XRD)

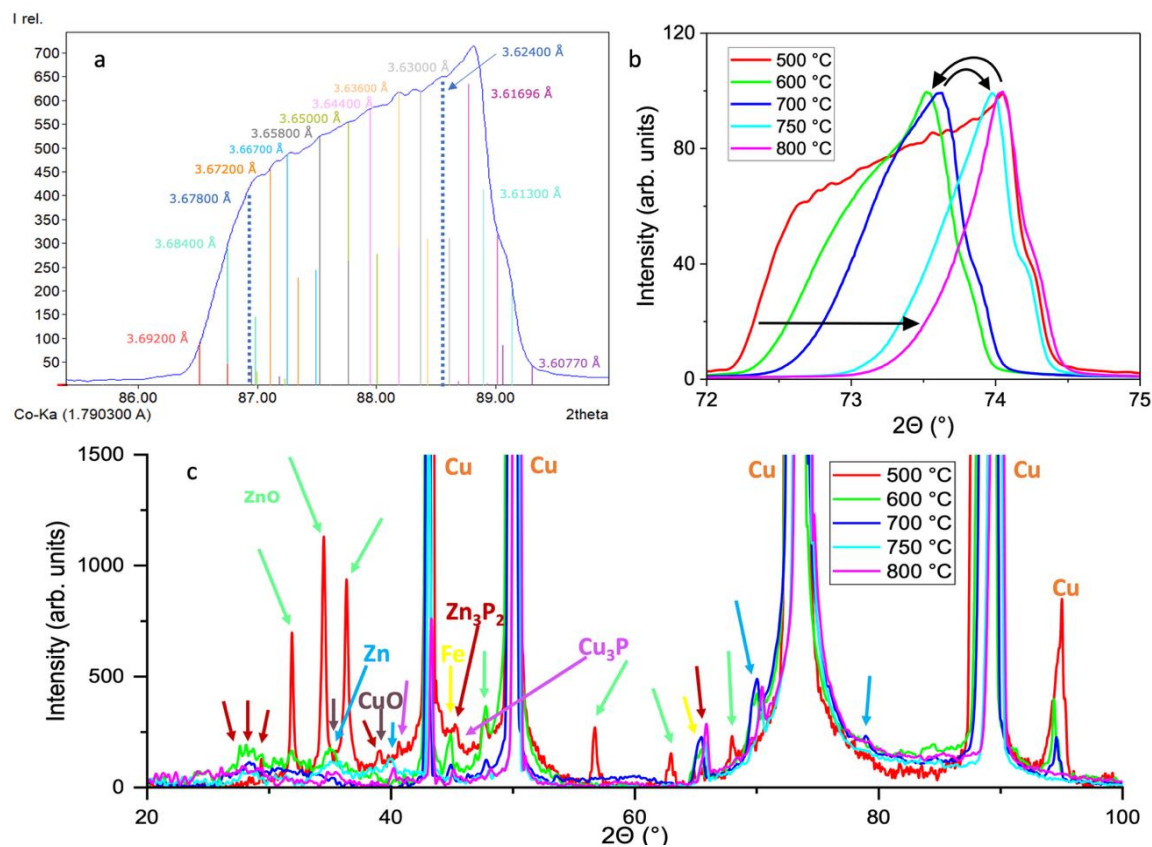

**Figure S5**. X-Ray Diffraction (XRD) results. a) XRD spectra of the brass alloy treated at 500 °C for 2h, focused on the (220) peak of the α-Cu phase. The colored markings present the referenced positions of different lattice spacing of the α-Cu determined from PDF database. The two blue dashed lines mark the calculated lattice spacing that were not available from the database. b) Presentation of XRD spectra of the brass alloy treated at different temperatures, focused on the (220) peak of the α-Cu phase. The dark arrows represent the change in the peak shape with increasing treatment temperature. c) XRD spectra of the brass alloy treated at different temperatures presented in an overlaid manner. The colored arrows represent the positions of the individual identified phases.

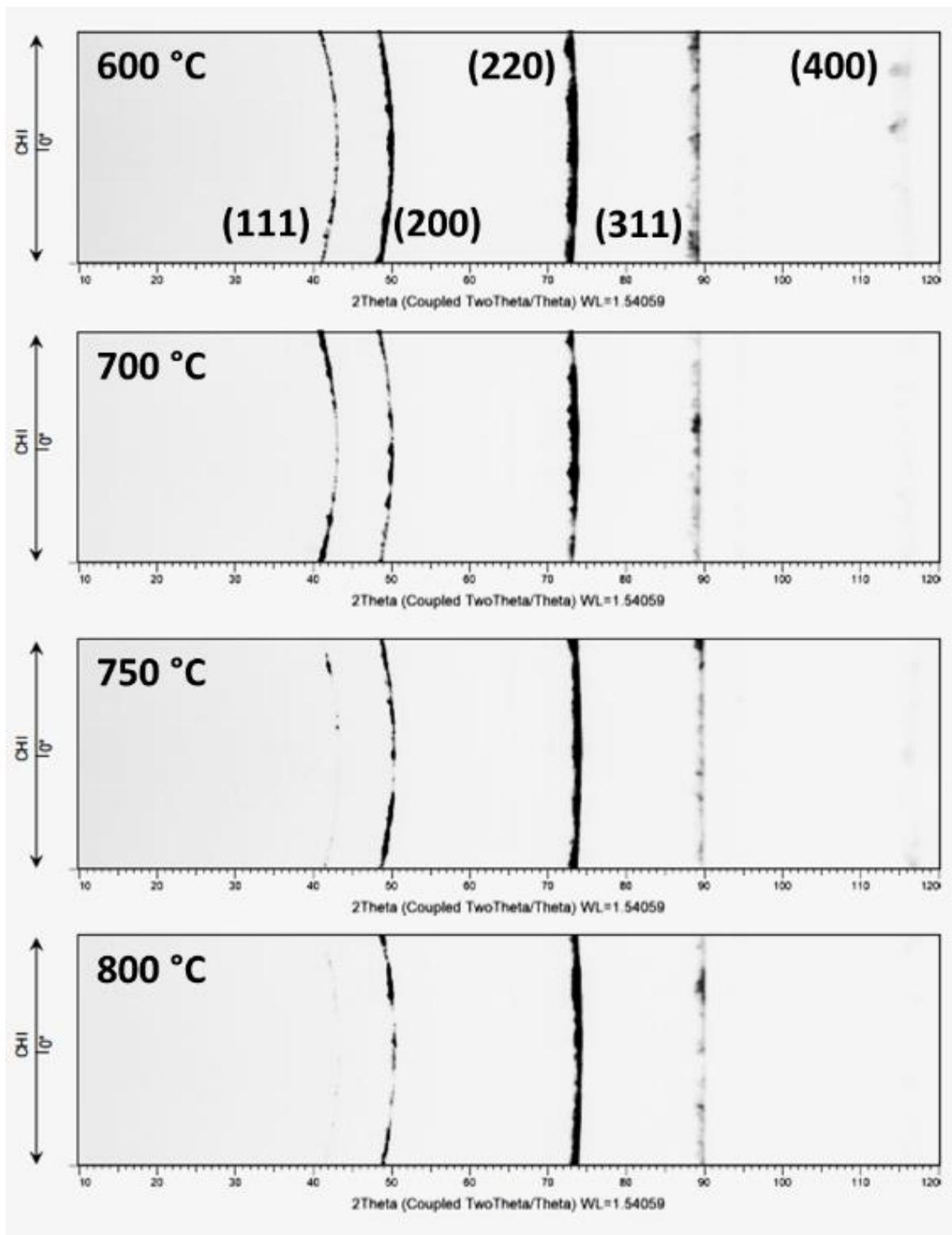

**Figure S6**. 2D images of the XRD dataset of the brass alloy treated at different temperatures.

From **Figure S5(a)**, the broadened (220) peak of the 500 °C dealloyed brass is presented that is associated to the α-Cu phase. The broadening is determined to be correlated to a series of lattice expansions reaching values up to 3.6692 Å. The broadening of the peak is considered to originate from the high distortion of the alloy at the surface and the inhomogeneous alloying of the α-Cu with Zn, confirming the Transmission Kikuchi Diffraction (TKD) results. This also leads to individual regions of more and less alloyed and distorted alloy, corroborated by the broad spotting in the diffraction rings from 2D image of the XRD data (see **Figure S6**) as well as by the TKD data. The broadening is registered for all α-Cu peaks, but focus is given onto the (220) peak due to the preferential texturing of the initial brass along [110]. **Figure S5(b)** shows that the high lattice values are disappearing with higher T. This coincides with coalescence of deformed material into pores by migration of defect (i.e. dynamic recovery and Ostwald ripening) with increasing treatment temperature, corroborating with nano-CT data. This is further seen by the immediate shift in the apex part of the broadened peak to higher diffraction angles (marked in **Figure S5(b)** by curved arrows on top) related to thermally induced homogenization of the alloy. With temperatures of 750°C and 800 °C there is a pronounced relaxation, determined from the shift of the peak's apex towards nominal values of pure Cu. This confirms the strong dealloying of Cu from Zn and recrystallization of the material (see **Figure S6**). This and Zn segregation is corroborated by the pronounced formation of 2 peaks at approximately 65° and 70°, related to $Zn_3P_2$ and Zn, respectively (see **Figure S5(c)**). Interestingly the Zn displays a high texture oriented with (110), indicating a crystallographic bounding of the Zn with the formed pores that could have a preferential formation with the texture direction of the Cu phase along [110]. The preferred recrystallization with the {220} of the α-Cu is clearly visible from the 2D XRD data, presented in **Figure S6**. The XRD also confirms the presence of ZnO, corroborating with the APT data on the Zn oxidation of the Zn-decorated pores.

Energy Dispersive X-Ray Spectroscopy (EDX) Cross-section Data

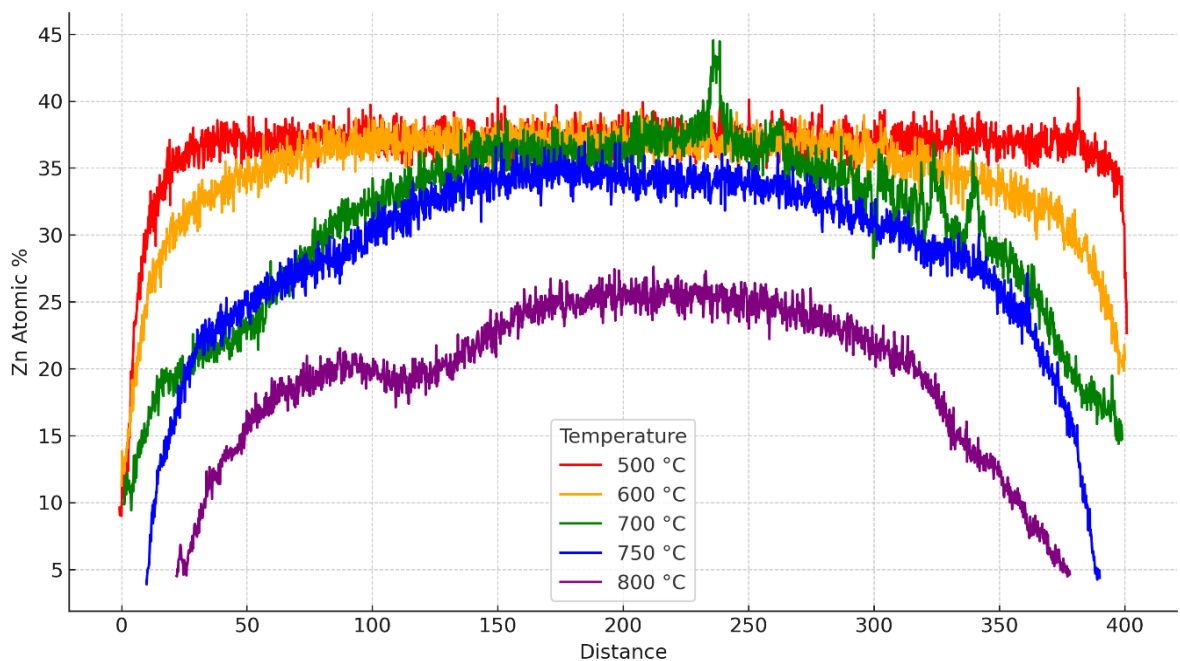

**Figure S7**. Zn 1-D concentration profile across embedded dealloyed samples prepared at different temperatures

Scanning Electron Microscope (SEM) EDX, X-Ray Photoelectron Spectroscopy (XPS), Transmission Electron Microscopy (TEM)-EDX, and Atom Probe Tomography (APT) measurements showing O concentrations

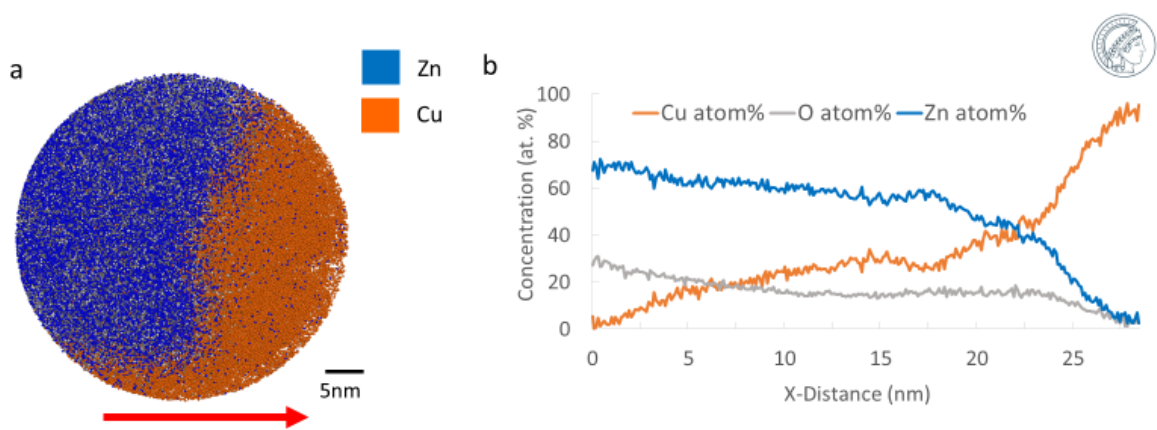

**Figure S8.** 500 °C / B500 surface APT data (a) X-Y view of B500 APT dataset from specific small pore (b) 1-D X-axis left-to-right elemental composition for Cu, Zn, and O illustrating high Zn gradient, direction noted by red arrow in (a)

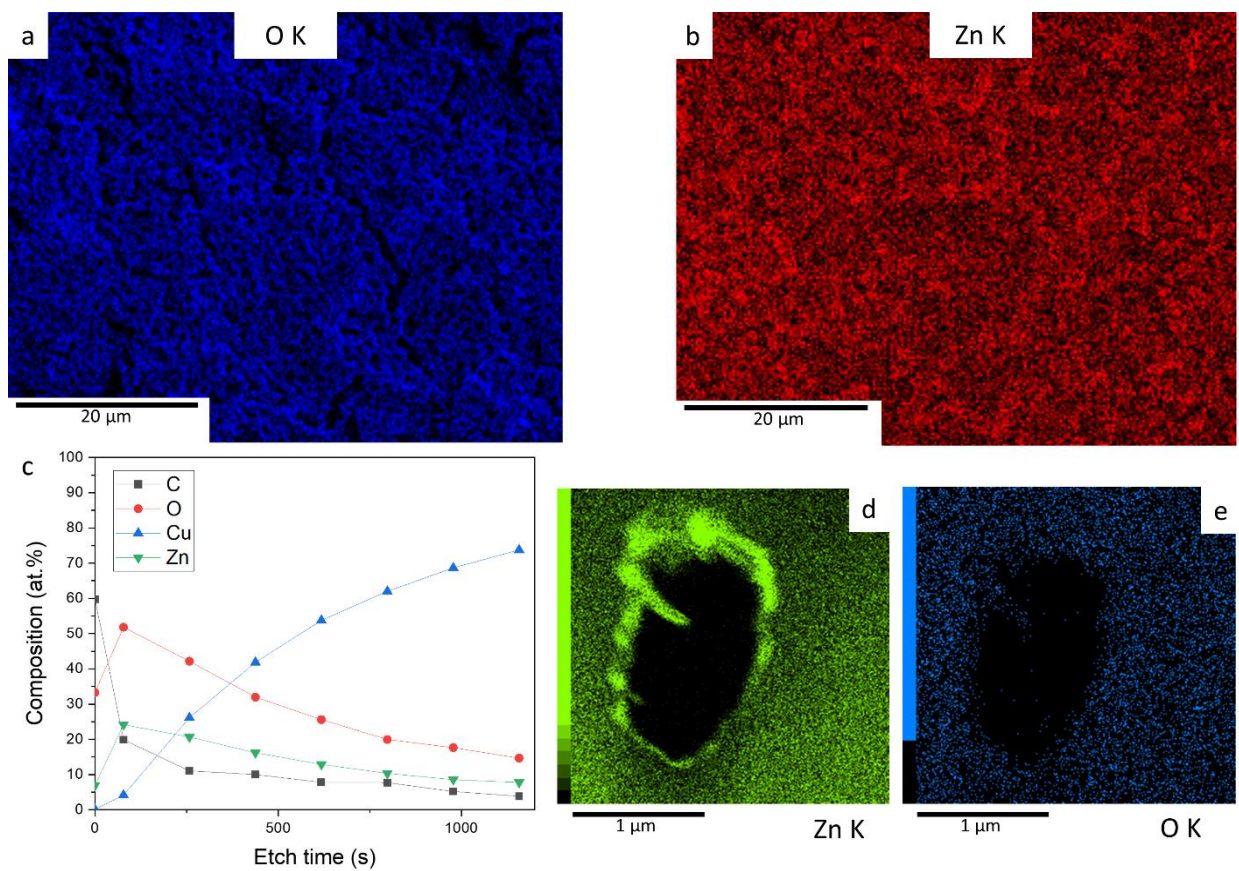

**Figure S9.** Supplemental data supporting ZnO. (a) O Kα EDX map, B500 surface (b) Zn Kα EDX map, B500 surface (c) XPS depth profile data for B500 to clean surface (d) Zn K EDX map, large pore on right, B500 (e) O K TEM EDX map, large pore on right, B500

## Nano-CT measurements

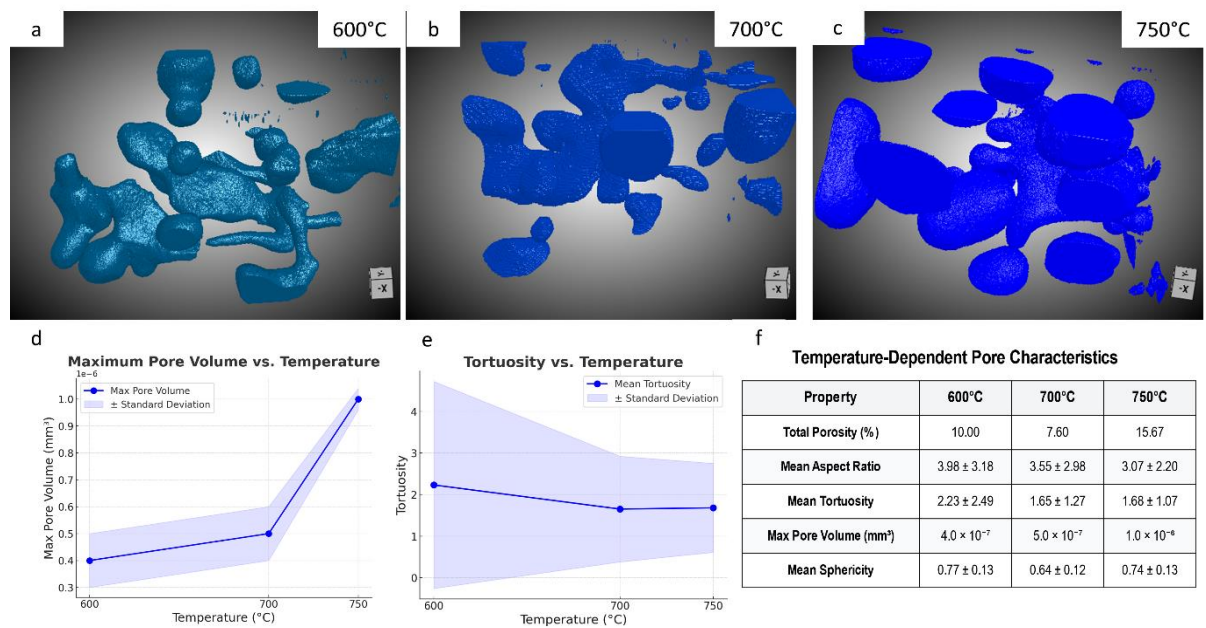

**Figure S10**. Nano-computed tomography (nano-CT) pore analysis for three temperature conditions (a) 600 °C (b) 700 °C (c) 750 °C (d) Maximum more volume – expected pore volume increase is observed with increasing temperature is observed with increasing temperature in the near-surface region, with standard deviation in light grey (e) Pores become more spherical with increasing temperature and become more spherical and less connected, as seen in decreasing tortuosity, with light blue indicating standard deviation (f) Pore statistics illustrating increased pore volume with temperature, showing coalescence

## Battery Performance Data

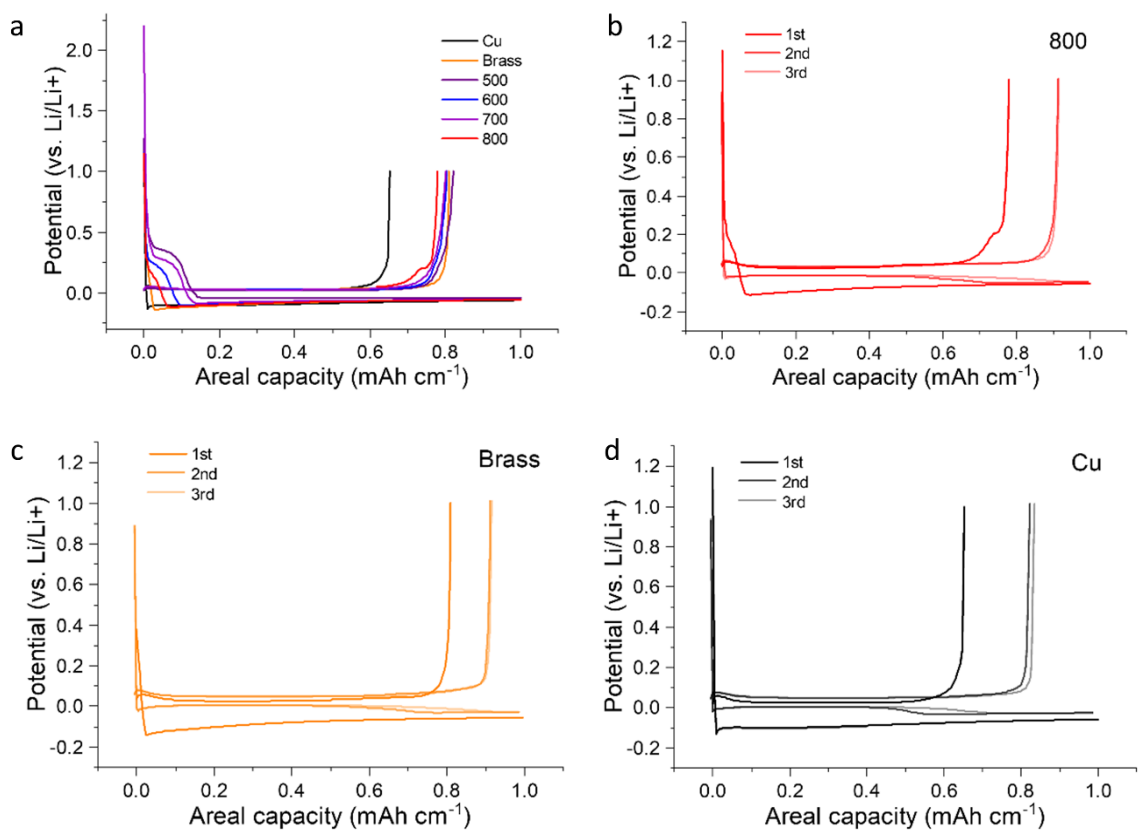

**Figure S11.** (a) Cathodic efficiency (CE) versus cycle number for the various cells, prepared with different collectors. Individual 1st – 3rd cycle efficiencies for (b) brass, (c) B800, and (d) Cu are shown individually.

## XPS Results

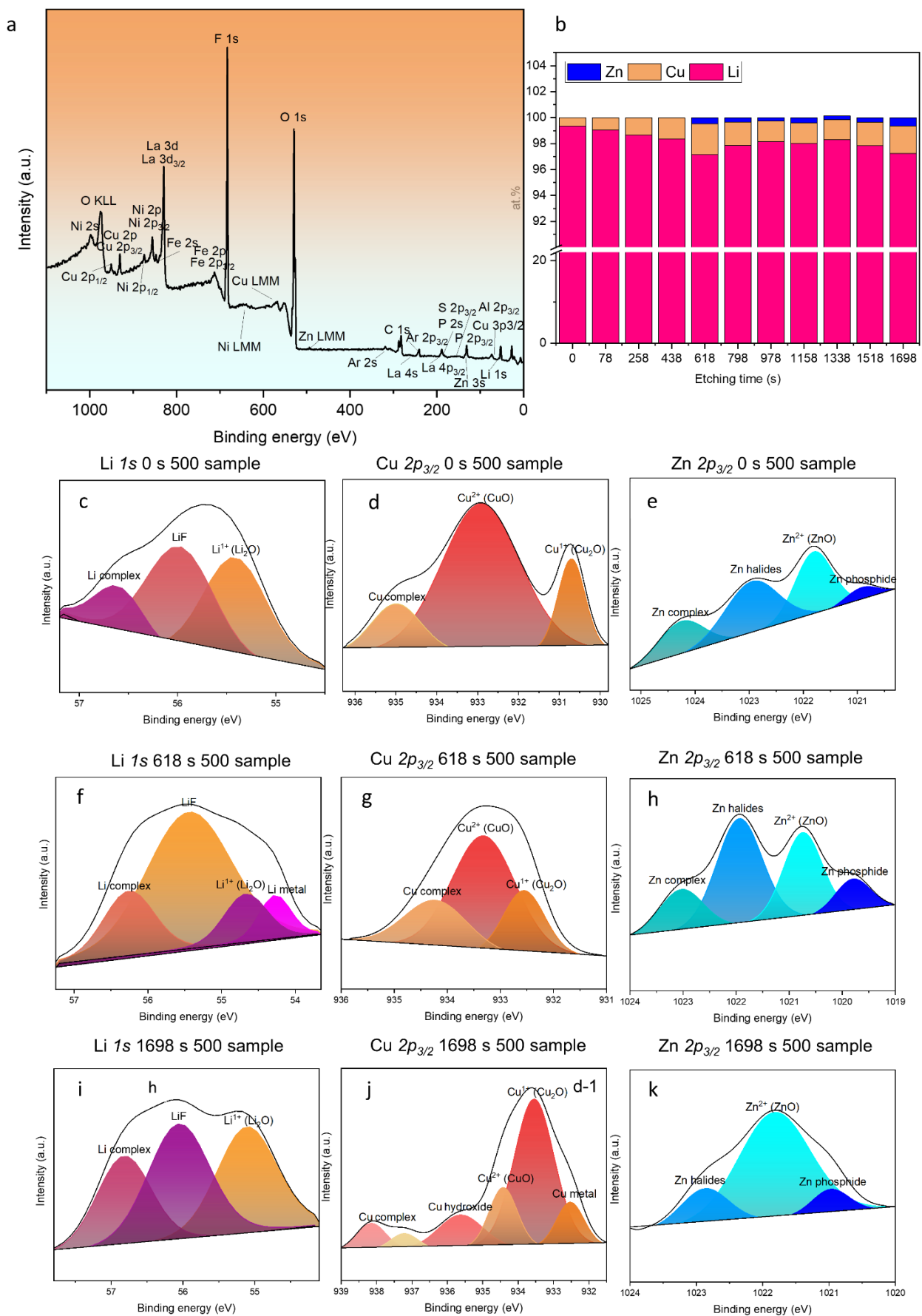

**Figure S12** XPS analysis results for 500 °C dealloyed brass (a) Survey scan results (b) composition of Zn, Cu, Li as a function of sputter etch time (c) Li high-resolution scan 0 s etch time (d) Cu hi-resolution scan 0 s etch time (e) Zn high-resolution scan 0 s etch time (f) Li high-resolution scan 618 s etch time (g) Cu high-resolution scan 618s etch time (h) Zn high-resolution scan 618s etch time (i) Li high-resolution scan 1698 s etch time (j) Cu high-resolution scan 1698 s etch time (k) Zn high resolution scan 1698 s etch time

The XPS survey results of B500 in **Figure S12** indicate the presence of the following elements O (*O 1s*), F (*F 1s*), C (*C 1s*), P (*P 2p*), S (*S 2p*), Li (*Li 1s*), Zn (*Zn 2p3/2*), La (*La $3d_{5/2}$*), Cu (*Cu $2p_{3/2}$*), Al (*Al 2p*), Ni (*Ni $2p_{3/2}$*) and Fe (*Fe $2p_{3/2}$*) (**Figure S12(a)**). The depth profile of the B500 sample (etching down to 1698 s at **Figure S12(b)** intervals) showed a decrease in the presence of *Li 1s* and an increase in *Cu $2p_{3/2}$* and *Zn $2p_{3/2}$*. Zn is less than 1 at.%, Cu 2.11 at.% and the rest is Li (~97 at.%). Furthermore, the high-resolution depth profile of Li 1s shows the presence of the $Li^{1+}$ oxidation state at 0 s of etching (**Figure S12(c)**), which correlates with the formation of $Li_2O$ (~55.45 eV), then the LiF (~55.95 eV) binding state, which correlates with the suspension used, and the Li complex binding state, which can be correlated with the LiFe/$LiFePO_4$ (~56.50 eV) (https://xpsdatabase.net/). At 618 s of etching (**Figure S(f)**), besides Li complex peak, LiF, $Li_2O$ peak, additional peak of Li metal (~54.35 eV) can be observed. The presence of $Li_2O$ is reduced and an increased presence of LiF and Li complex can be observed, which correlates with the removal of Li oxides on the surface. At the last etching step, 1698 s (**Figure S12(i)**), the presence of three binding states can be observed, $Li_2O$, LiF and Li complex. Where all three binding states have similar presence. The observation of *Cu $2p_{3/2}$* in the 500 °C dealloyed sample, at 0 s (**Figure S12(d)**), shows the presence of $Cu^{1+}$ oxidation state (~930.68 eV), which is proposed to be in $Cu_2O$ binding state, $Cu^{2+}$ (~932.81 eV), which is suggested to be in CuO binding state and Cu complex peak (~935.12 eV), which is dedicated to the combined peak of Cu hydroxide and Cu sulphate. At 618 s (**Figure S12(g)**), the binding states of Cu, are $Cu^{1+}$, $Cu^{2+}$ and Cu complex, which is similar observation as at the previous etching time. At the last etching time (1698 s) (**Figure S12(j)**), the presence of 5 binding state is observed. Cu metal (~932.58 eV), $Cu^{1+}$ (~933.21 eV), $Cu^{2+}$ (~934.25 eV), Cu hydroxide (~935.61 eV) and two peaks of Cu complex state (~937.28 eV and ~938.24 eV), which are dedicated to the complex sulphates and $Cu^{1+}$ combined with other oxides. The observations of *Zn $2p_{3/2}$* peak at 0 s (**Figure S12(e)**), showed the presence of the following binding state of Zn: Zn phosphides (~1020.82 eV), oxidation state of $Zn^{2+}$ as in ZnO (~1021.78 eV), Zn halides (~1022.88 eV) and complex peak of Zn (~1024.18 eV), which can be dedicated to complex Zn sulphate or even complex bound water/hydroxides. The next observation at 618 s (**Figure S12(h)**), shows the presence of the same binding states, which are in average shifted by 0.55 eV. At 1698 s (**Figure S12(k)**), there are predominant 3 binding states of Zn: Zn phosphides (~1020.97 eV), $Zn^{2+}$ ($Zn_2O$)(~1021.88 eV) and Zn halides (~1022.87 eV).

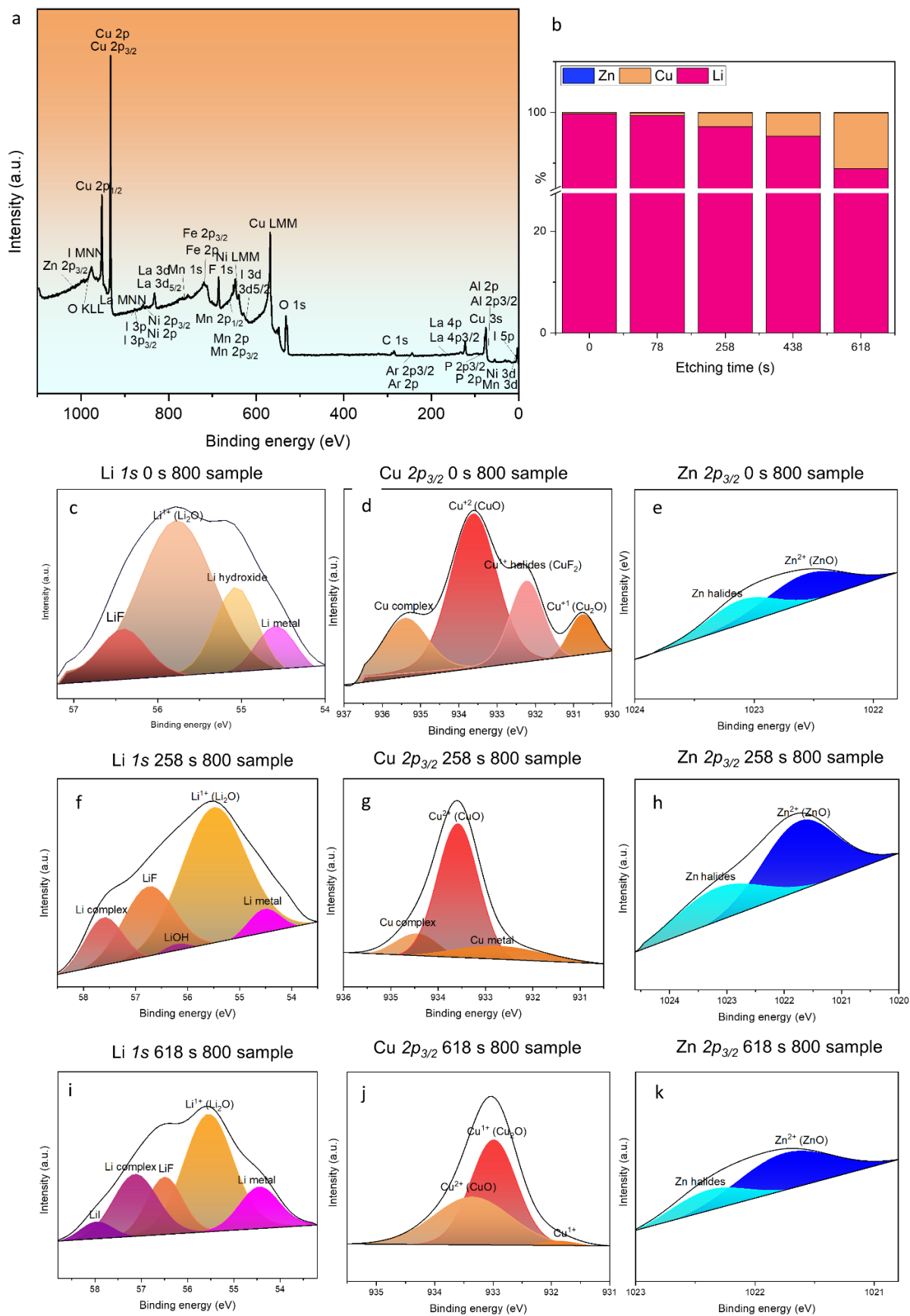

**Figure S13.** XPS results 800 °C sample. (a) Survey scan results (b) composition of Zn, Cu, Li as a function of sputter etch time (c) Li high-resolution scan 0 s etch time (d) Cu hi-resolution scan 0 s etch time (e)

Zn high-resolution scan 0 s etch time (f) Li high- resolution scan 258 s etch time (g) Cu high-resolution scan 250 s etch time (h) Zn high-resolution scan 250 s etch time (i) Li high-resolution scan 618 s etch time (j) Cu high-resolution scan 618 s etch time (k) Zn high resolution scan 618 s etch time

The XPS results of sample B800 (**Figure S13**) indicate the presence of the following elements: O (*O 1s*), F (*F 1s*), C (*C 1s*), P (*P 2p*), I (*I 3p$_{3/2}$*), Li (*Li 1s*), Zn (*Zn 2p$_{3/2}$*), La (*La 3d$_{5/2}$*), Cu (*Cu 2p$_{3/2}$*), Al (*Al 2p*), Ni (*Ni 2p$_{3/2}$*) and Fe (*Fe 2p$_{3/2}$*) (SM a). The depth profile of B800 (etching down to 618 s at **Figure S13(b)** intervals) showed a decrease in the presence of *Li 1s* and an increase in *Cu 2p$_{3/2}$* and *Zn 2p$_{3/2}$*. Zn is less than 1 at.%, Cu 11 at.% and the rest is Li (~89 at.%). The detail analysis showed for in-depth profile of *Li 1s* at 0 s (**Figure S13(c)**) shows the presence of the Li metal binding state (~54.55 eV), Li in hydroxide binding state (~55.09 eV), Li$^{1+}$ in Li$_2$O state (~55.78 eV) and Li complex peak, which can be correlated to combination of LiF/ LiFe/LiFePO$_4$ (~56.48 eV). At 258 s of etching (**Figure S13(e)**), the presence of Li metal (~54.78 eV), Li$^{1+}$ as Li$_2$O (~55.47 eV), LiOH (~56.17 eV), LiF (~56.67 eV) and Li complex state (~57.67 eV), which can be correlated to the halides (combination of F/I) and phosphates. At the last etching step, 618 s (**Figure S13(i)**), the presence of five binding states can be observed: Li metal (~54.39 eV), Li$^{1+}$ as in Li$_2$O (~55.59 eV), LiF (~56.49 eV), Li complex (~57.09 eV) and LiI (~58 eV) https://xpsdatabase.net/. Li complex state is associated with phosphates, which are correlated to the presence of P in the sample. The observation of *Cu 2p$_{3/2}$* in 800 sample, at 0 s (**Figure S13(d)**), shows the presence of Cu$^{1+}$ oxidation state (~930.68 eV), which is proposed to be in Cu$_2$O binding state, Cu$^{1+}$ (~932.26 eV), which is suggested to be in CuF binding state, Cu2+ as in CuO (~933.66 eV) and Cu complex (~935.26 eV), which is dedicated to the combined peak of Cu hydroxide and Cu sulphate. At 258 s (**Figure S13(g)**), only three binding states of Cu are detected: Cu metal (~932.70 eV), Cu$^{2+}$ as in CuO (~933.65 eV) and Cu complex (934.45), which is similar observation as at the previous etching time. At the last etching time (618 s, **Figure S13(j)**), the presence of 3 binding state is observed: Cu$^{1+}$ (~931.82 eV), Cu$^{1+}$ (~933.02 eV) as Cu$_2$O and Cu$^{2+}$ (~933.32.25 eV) as CuO. The observations of *Zn 2p$_{3/2}$* peak at 0 s in 800 sample (**Figure S13(e)**), showed the presence of the following two binding state of Zn: Zn$^{2+}$ as ZnO (~1022.60 eV) and Zn halides (~1023.12 eV). The observations of *Zn 2p$_{3/2}$* peak at 258 s (**Figure S13(h)**) in 800 sample, showed the presence of the following two binding state of Zn: Zn$^{2+}$ as ZnO (~1021.65 eV) and Zn halides (~1023.15 eV). At the last etch step 618 s (**Figure S13(k)**), there are predominant 2 binding states of Zn: Zn$^{2+}$ as ZnO (~1021.74 eV) and Zn halides (~1022.39 eV).

### Thermocalc and DICTRA simulations

An example equilibrium phase diagram for the Cu-Zn system is shown in **Figure S14**, with the tested temperatures marked with '*'. The Cu-Zn phase diagram for vacuum generated through Thermo-Calc simulations, **Figure S15(a)**, maps the equilibrium phases (α-FCC, β-BCC, liquid, and gas) as functions of temperature and pressure, while **Figure S15(b)** illustrates the thermodynamic driving force for Zn vaporization across the temperature range studied. When simulating Zn concentration profiles at $1 \times 10^{-5}$ mbar vacuum, the DICTRA models predict no measurable Zn depletion in the bulk material at both 500 °C and 600 °C, with compositional changes restricted to the near-surface regions. For the Cu-Zn system, the difference between at.% and wt.% is less than 1% when converting between them, so they are used interchangeably here. Even at 800 °C, the simulations suggest only modest bulk depletion, predicting a central Zn concentration of approximately 34 wt.%.

The comparison between simulations and experiments reveals significant temperature-dependent deviations. For the specific case of the Cu-Zn system, atomic and weight % are interchangeable within 1%. At 500 °C in **Figure** (**a**), the experimental concentration profiles align closely with simulation predictions in both gradient and absolute values. However, at 800 °C in **Figure (b)**, experimental measurements show substantially greater Zn depletion than predicted, with EDX-measured bulk Zn composition reaching approximately 25 wt.% compared to the simulated 34 wt.%. Full plots of all temperature 1-D profiles can be found in **Figure S7** compared with simulated DICTRA / Thermocalc results in **Figure S,** which show similar increasing deviations with increasing temperature. The volume fraction phase diagram is shown in **Figure** and temperature versus volume fraction for alpha, beta, and liquid phases in **Figure S19**.

These differing transport mechanisms help explain the increasing divergence between experimental results and simulations at higher temperatures. In lower temperature regimes, the primary transport mechanism is planar surface diffusion (iv), which the DICTRA simulations model well, so the simulations and experimental data closely match for the alpha phase of brass. As temperatures increase above 700 °C, the enhanced bulk diffusion (iii) transport mechanism increases, while the evaporation-condensation (iv) becomes more efficient within deeper pores, so that multiple mechanisms are simultaneously operating. Additionally, the DICTRA simulations only consider transport in the alpha phase, where the phase diagram shows the beta phase of brass becomes important as the system transitions into a mixed state. The omission of brass beta phase transport mechanisms, which likely have different diffusion coefficients and transport properties, explains why the simulations significantly underestimate Zn depletion from the bulk material at higher temperatures.

Since the DICTRA simulations only consider transport in the alpha phase, where the phase diagram shows the beta phase of brass becomes important as the system transitions into a mixed state. The omission of brass beta phase transport mechanisms, which likely have different diffusion coefficients and transport properties, explains why the simulations significantly underestimate Zn depletion from the bulk material at higher temperatures.

This increasing deviation between simulation and experimental results at elevated temperatures suggests that the theoretical models may not fully capture all temperature-activated mechanisms of Zn transport. While the phase diagram in **Figure S15(a)** and driving force calculations in **Figure S15(b)** correctly predict the thermodynamic tendency for Zn volatilization under vacuum ($1 \times 10^{-5}$ mbar), particularly at higher temperatures, the experimental data reveals that actual dezincification proceeds more extensively than thermodynamic and kinetic simulations would indicate. The surface evolution during dealloying provides important insights into the mechanism: at 500 °C, extensive surface porosity develops (covering approximately 50% of the surface area), while at 800 °C, fewer but deeper pores form (approximately 10% surface coverage with larger subsurface volumes). This temperature-dependent pore morphology suggests that at higher temperatures, the system favors fewer, more penetrating dezincification pathways rather than widespread surface porosity. These deep, interconnected pore networks likely facilitate more efficient bulk Zn transport to the surface than predicted by the simulations, which primarily consider planar surface evaporation.

At 900 °C, DICTRA simulations predict a retention of 27 wt.% Zn in the bulk, but an experimental sample underwent complete dezincification. Since there was severe physical warping that resembled crumpled foil, standard surface and cross-sectional analyses were not feasible for the 900 °C sample. This extreme deformation is consistent with the proximity to the partially liquid phase region of the Cu-Zn system above 900 °C. However, micro-capillary XRD performed on the mid-section of a through-cut specimen confirmed complete transformation to pure Cu, as evidenced by lattice parameters matching those of pure Cu with no detectable Zn-containing phases – XRD data show in **Figure S20**.

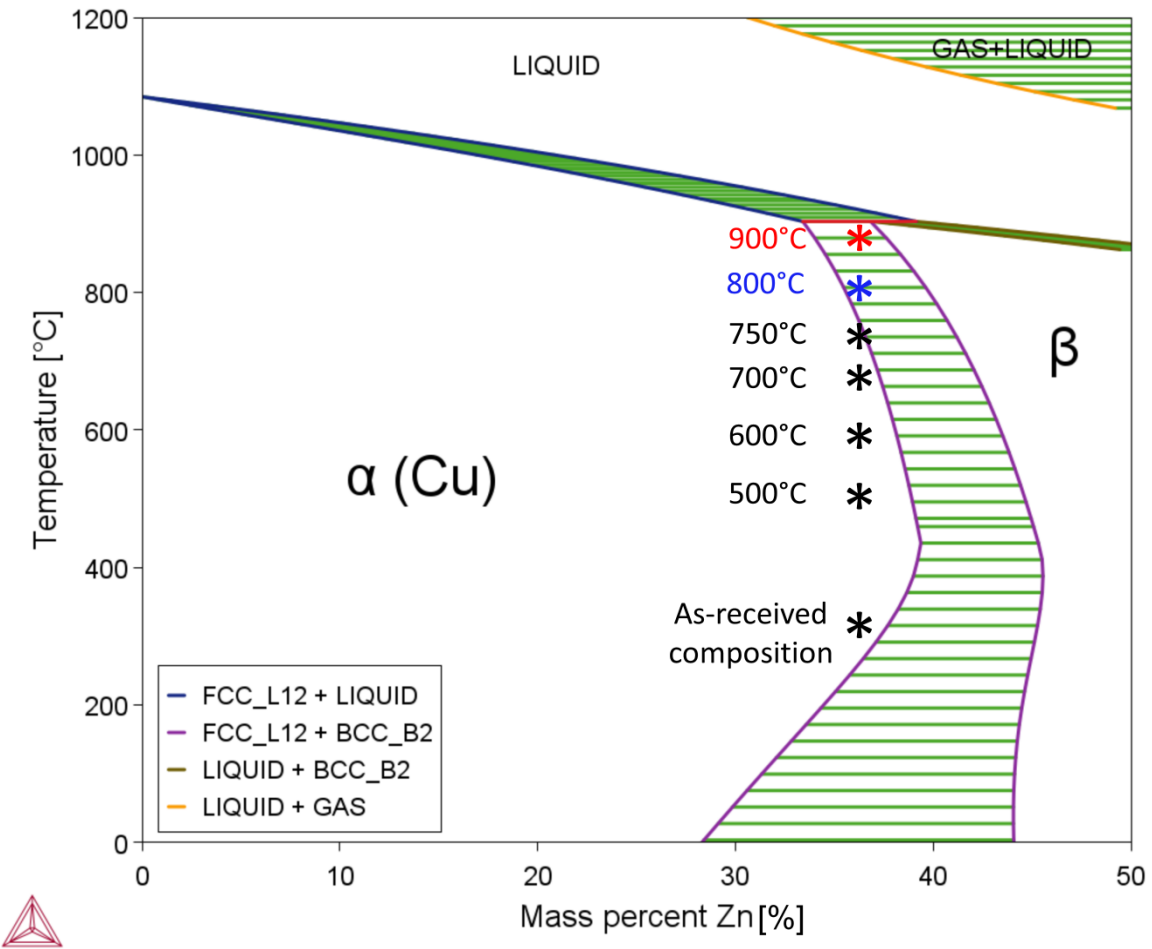

**Figure S14.** Equilibrium Thermocalc Cu-Zn phase diagram, as denoted by the black asterisks; the optimal temperature of 800 °C is denoted by the blue asterisk. Temperature of 900 °C effectively destroyed the specimen, as noted by the red asterisk

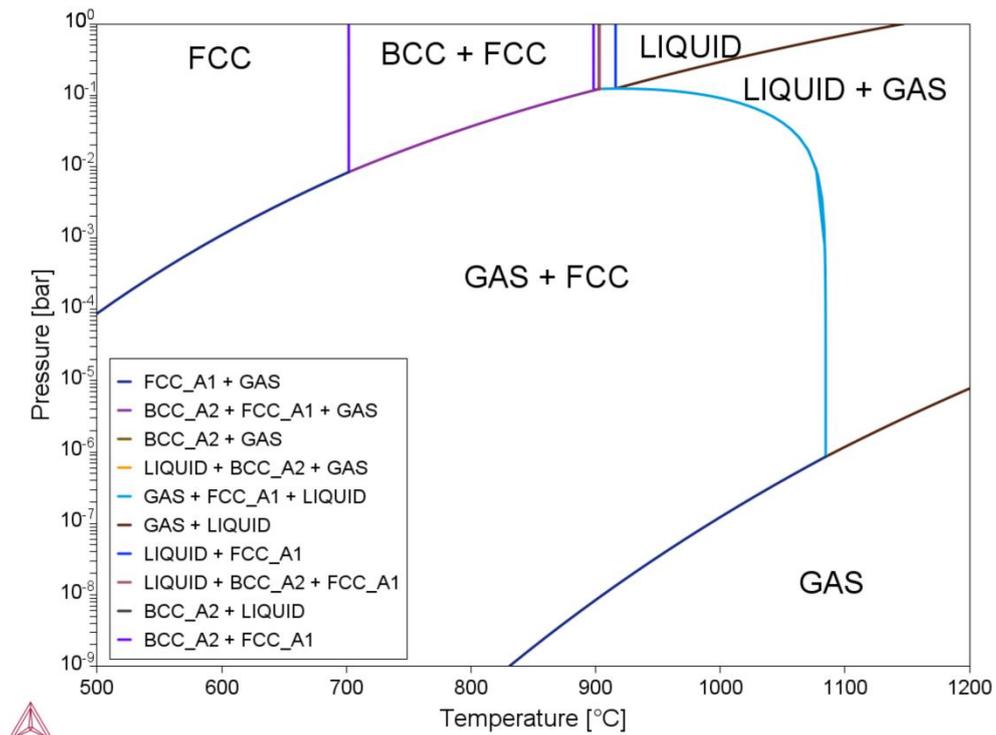

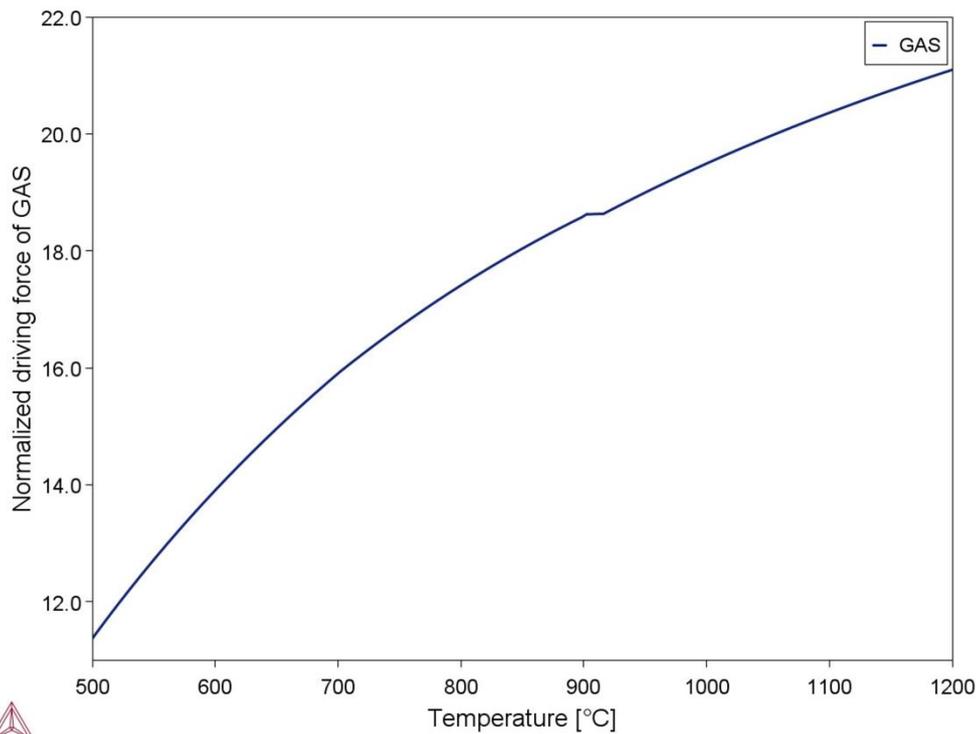

**Figure S15.** (a) Thermocalc simulation of the phase diagram for the Cu-Zn binary alloy system under varying vacuum pressures, showing the stability of different phases (e.g. the FCC α-phase and BCC Cu-rich β-phase as in Figure 1, liquid, and gas) as functions of temperature and pressure (b) Thermocalc simulation of the normalized driving force for Zn volatilization from the bulk Cu-Zn alloy as a function of temperature, calculated at a vacuum pressure of $10^{-5}$ bar

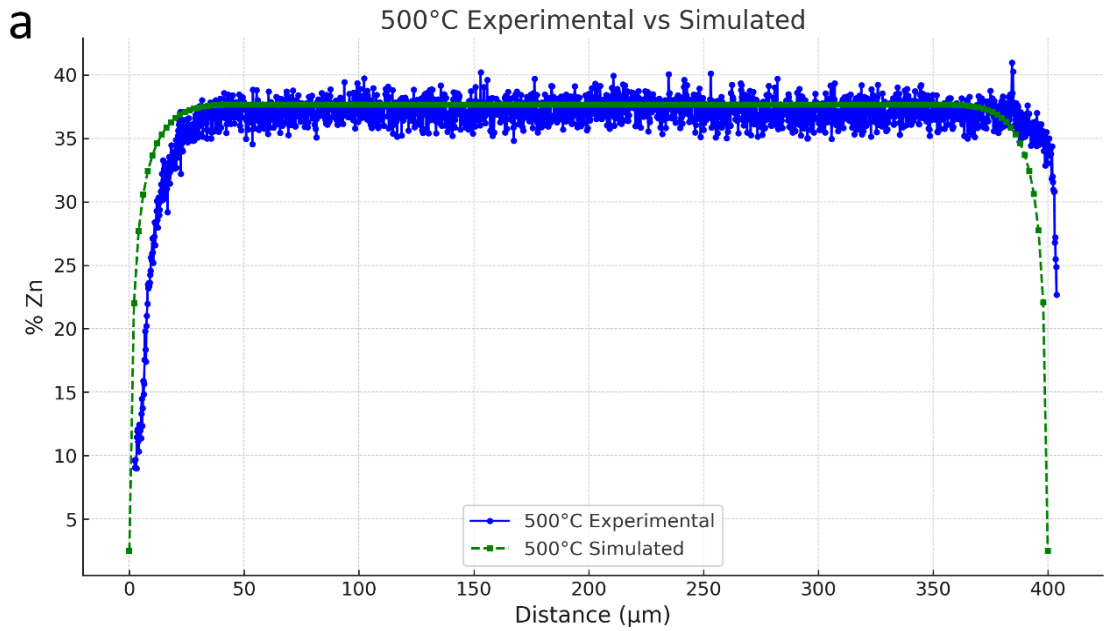
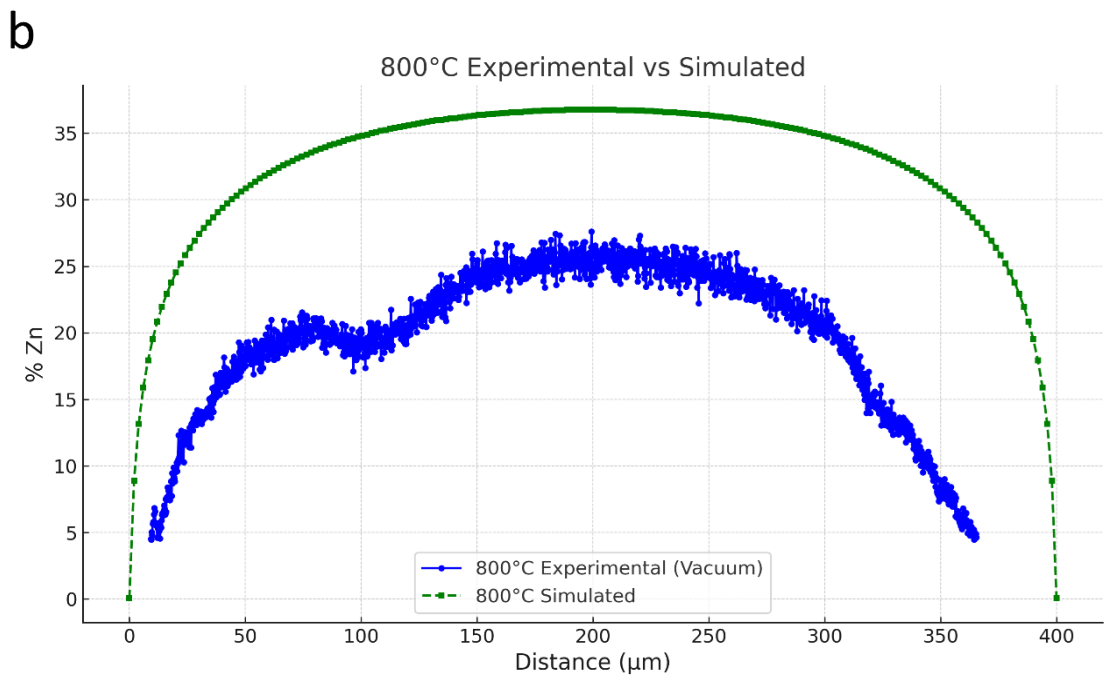

**Figure S16.** Comparison of experimental and simulated Zn cross-sectional profiles at (a) 500 °C and (b) 800° C

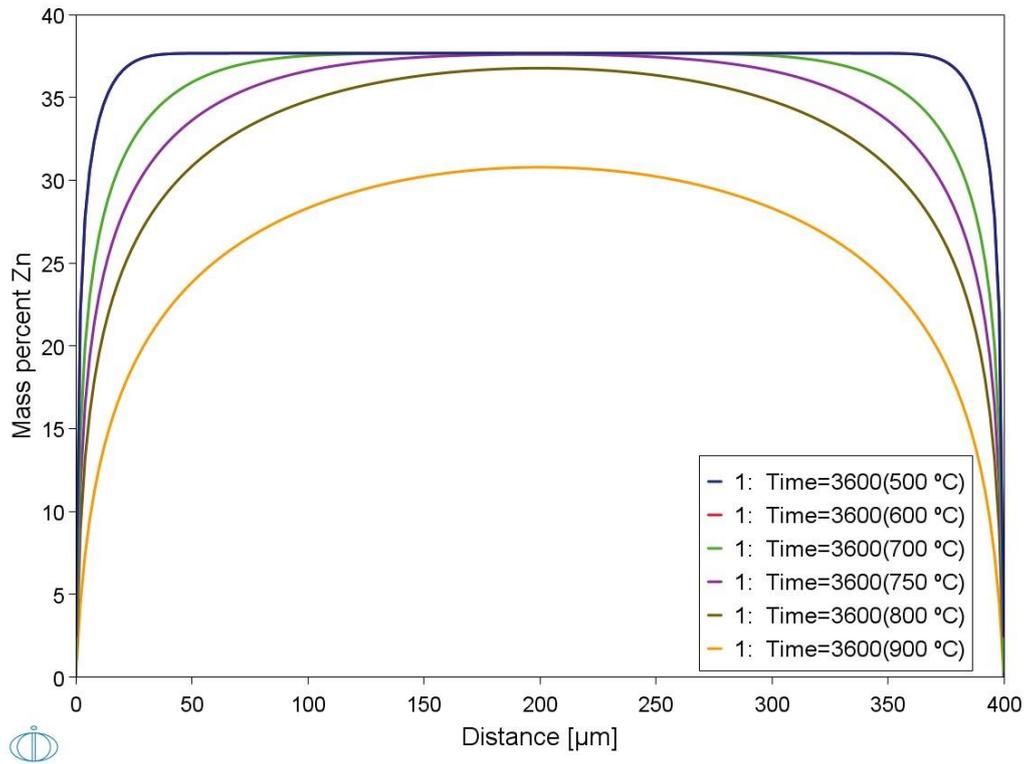

**Figure S17**. Thermocalc & Dictra simulations of expected 1-D Zn profile

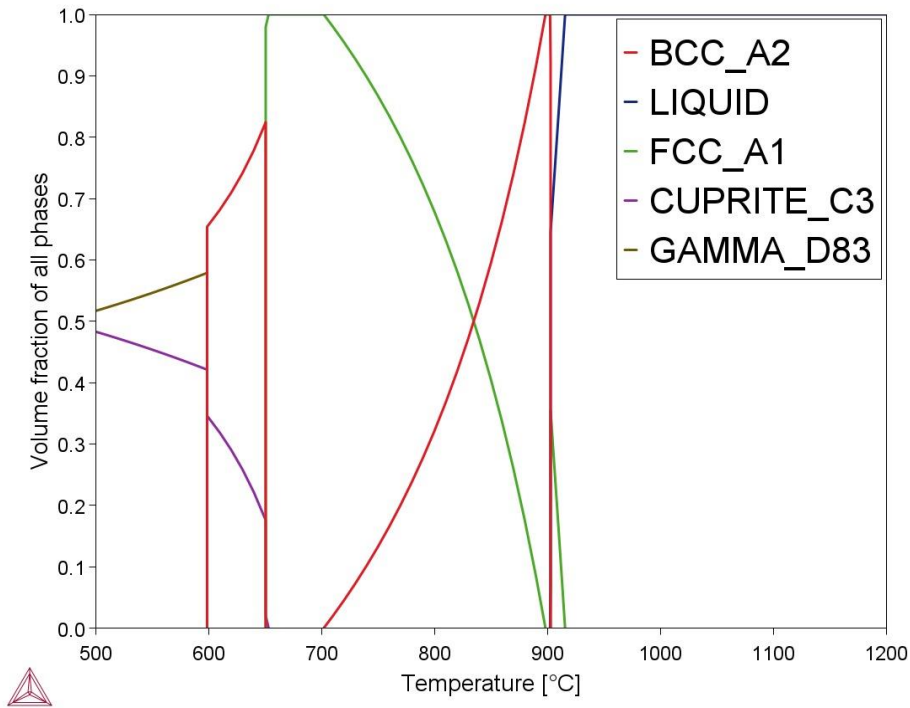

**Figure S18.** Thermocalc simulation of phase diagram for volume.

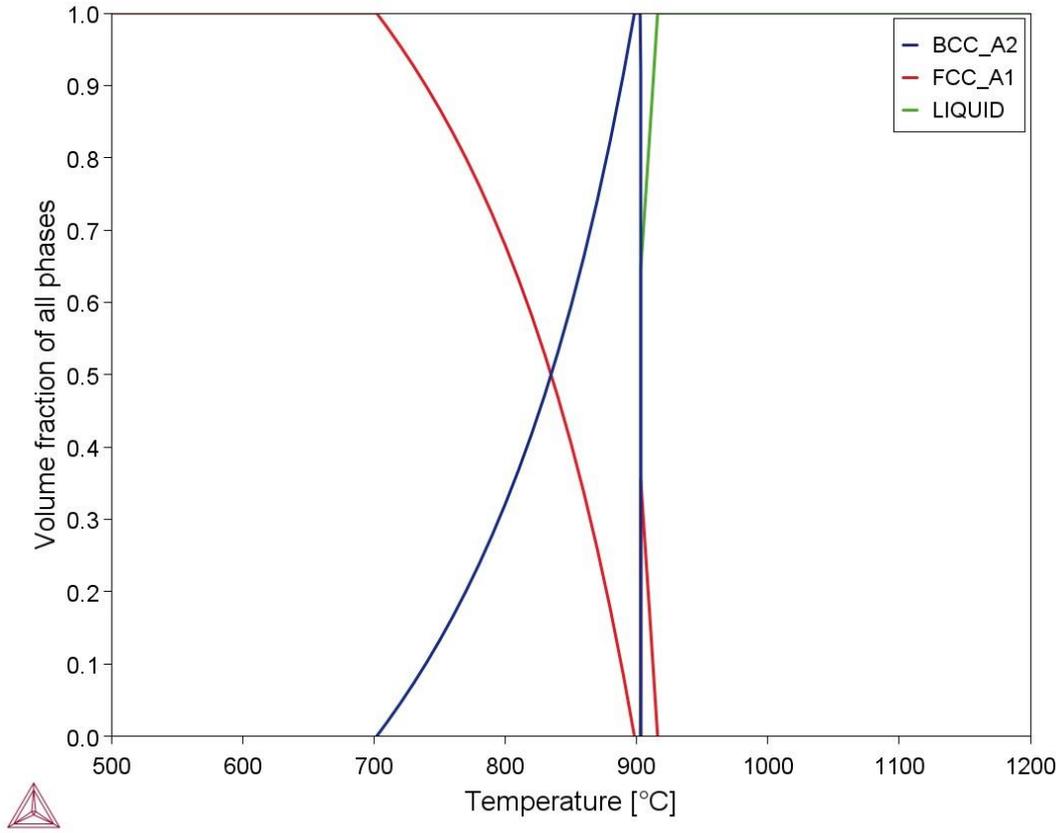

**Figure S19.** Thermocalc simulation showing temperature versus volume phase fraction for alpha-beta-liquid states.

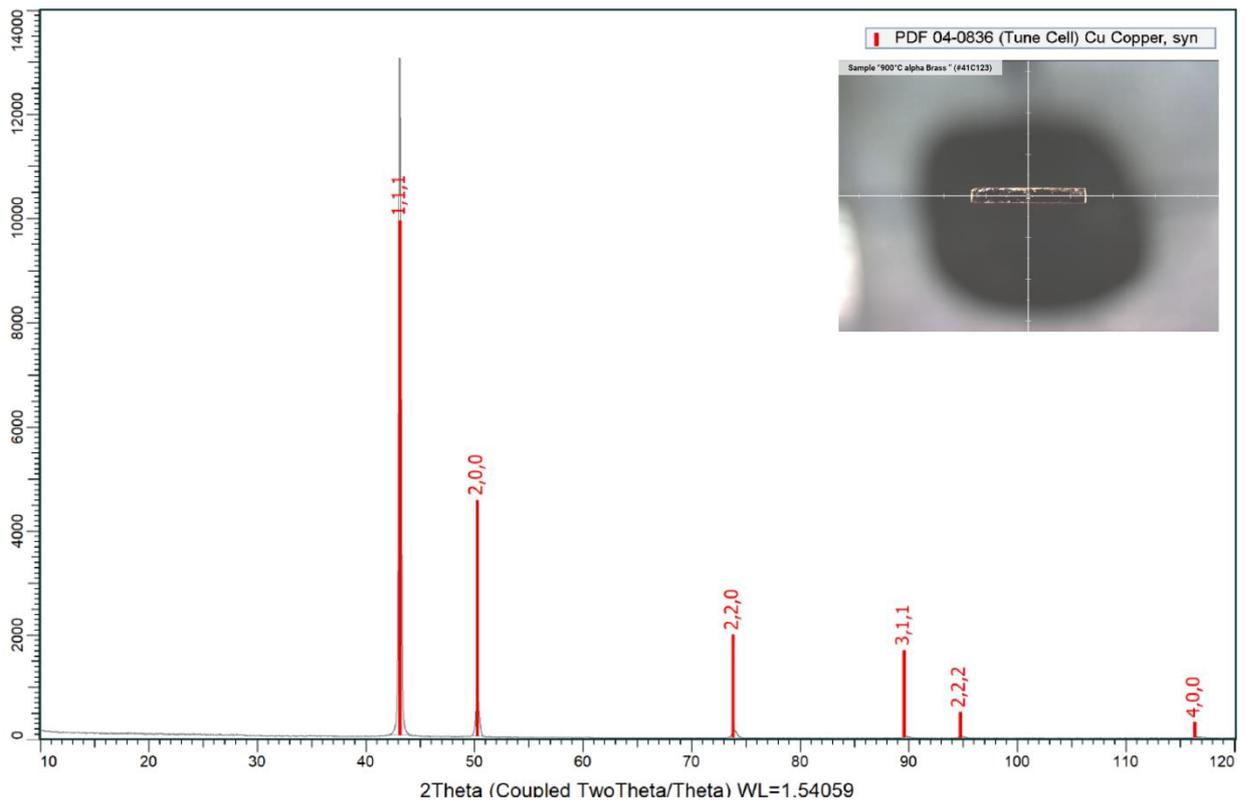

**Figure S20.** XRD data with inset picture of 900C sample

## Supplementary Instrument Settings

### XRD measurement configurations

| Measurement Condition | Instrument | Source Type | Primary Optic | Geometry |
|---|---|---|---|---|
| 500°C Bulk XRD (2 hours) | Bruker D8 Advance A25-X1 | Co Kα | Slit optic / Soller slit 4.0° | Bragg-Brentano / symmetrical |
| 500°C Grazing Incidence XRD | Rigaku Smartlab 9kW | Cu Kα | Goebel mirror, Soller slit 5.0° | Parallel Beam / GIXRD |
| 600°C Bulk XRD | Rigaku Smartlab 9kW | Cu Kα | Micro Area | Parallel beam / symmetrical |
| 700°C Bulk XRD | Rigaku Smartlab 9kW | Cu Kα | Micro Area | Parallel beam / symmetrical |
| 750°C Bulk XRD | Rigaku Smartlab 9kW | Cu Kα | Micro Area | Parallel beam / symmetrical |
| 800°C Bulk XRD | Rigaku Smartlab 9kW | Cu Kα | Micro Area | Parallel beam / symmetrical |
| 900°C Microcapillary XRD | Rigaku Smartlab 9kW | Cu Kα | CBOμ (Diameter 125μm) | Parallel beam / symmetrical |

**Table T1.** XRD instrument configuration

| Measurement Condition | Secondary Optic | Detector Type | Power Setting | Scan Speed / 2Θ Increment |
|---|---|---|---|---|
| 500°C Bulk XRD (2 hours) | Soller slit 4.0° | LYNXEYE XE-T (Energy dispersive 1D) | 35 kV, 40 mA | 225s / step, 0.009° |
| 500°C Grazing Incidence XRD | Soller slit PSA90 0.228° | HyPix 3000 (2D) | 45 kV, 200 mA | 0.75°/min, 0.01° |
| 600°C Bulk XRD | N/A | HyPix 3000 (2D) | 45 kV, 200 mA | 1°/min, 0.01° |
| 700°C Bulk XRD | N/A | HyPix 3000 (2D) | 45 kV, 200 mA | 1°/min, 0.01° |
| 750°C Bulk XRD | N/A | HyPix 3000 (2D) | 45 kV, 200 mA | 1°/min, 0.01° |
| 800°C Bulk XRD | N/A | HyPix 3000 (2D) | 45 kV, 200 mA | 1°/min, 0.01° |
| 900°C Microcapillary XRD | N/A | HyPix 3000 (2D) | 45 kV, 200 mA | 1°/min, 0.01° |

**Table T2.** Additional XRD instrument configuration details